\documentclass[aps,pre,reprint,superscriptaddress,10pt]{revtex4-1}
\usepackage{amsmath, amssymb, graphicx, color, braket}

\begin{document}

\title{Unified theory of quantum phase transitions in quantum dots with gapped host bands}

\author{Peter Zalom}
\email{zalomp@fzu.cz}
\affiliation{Institute of Physics, Czech Academy of Sciences, Na Slovance 2, CZ-18221 Praha 8, Czech Republic}

\author{Martin \v{Z}onda}
\email{martin.zonda@karlov.mff.cuni.cz}
\affiliation{Department of Condensed Matter Physics, Faculty of Mathematics and Physics, Charles University, Ke Karlovu 5, CZ-12116  Praha 2, Czech Republic}

\date{\today}

\begin{abstract}
We present a unified theory of quantum phase transitions for half-filled quantum dots (QDs) coupled to gapped host bands. We augment the bands by additional weakly coupled metallic lead which allows us to analyze the system by using standard numerical renormalization group techniques. The ground state properties of the systems without the additional metallic lead are then extrapolated in a controlled way from the broadened subgap spectral functions. We show that a broad class of narrow-gap-semiconductor tunneling densities of
states (TDOSs) support the existence of two distinct phases known from their superconducting counterpart. Namely,
$0$ phase which is marked by the singlet ground state and the $\pi$ phase regime with the doublet ground state.  To keep a close analogy with the superconducting case, we focus on the influence of particle-hole asymmetry of the TDOS of the subgap spectral features. Nevertheless, we also discus the possibility of inducing singlet-doublet quantum phase transitions in experimental setups by varying the filling of the QD. In addition, for gapped TDOS functions with smoothed gap edges, we demonstrate that all subgap peaks may leak out of the gap into the continuous part of the spectrum, an effect which has no counterpart in the superconducting Anderson model.
\end{abstract}


\maketitle

\section{Introduction \label{sec_intro}}

\begin{figure*}
	\includegraphics[width=2.0\columnwidth]{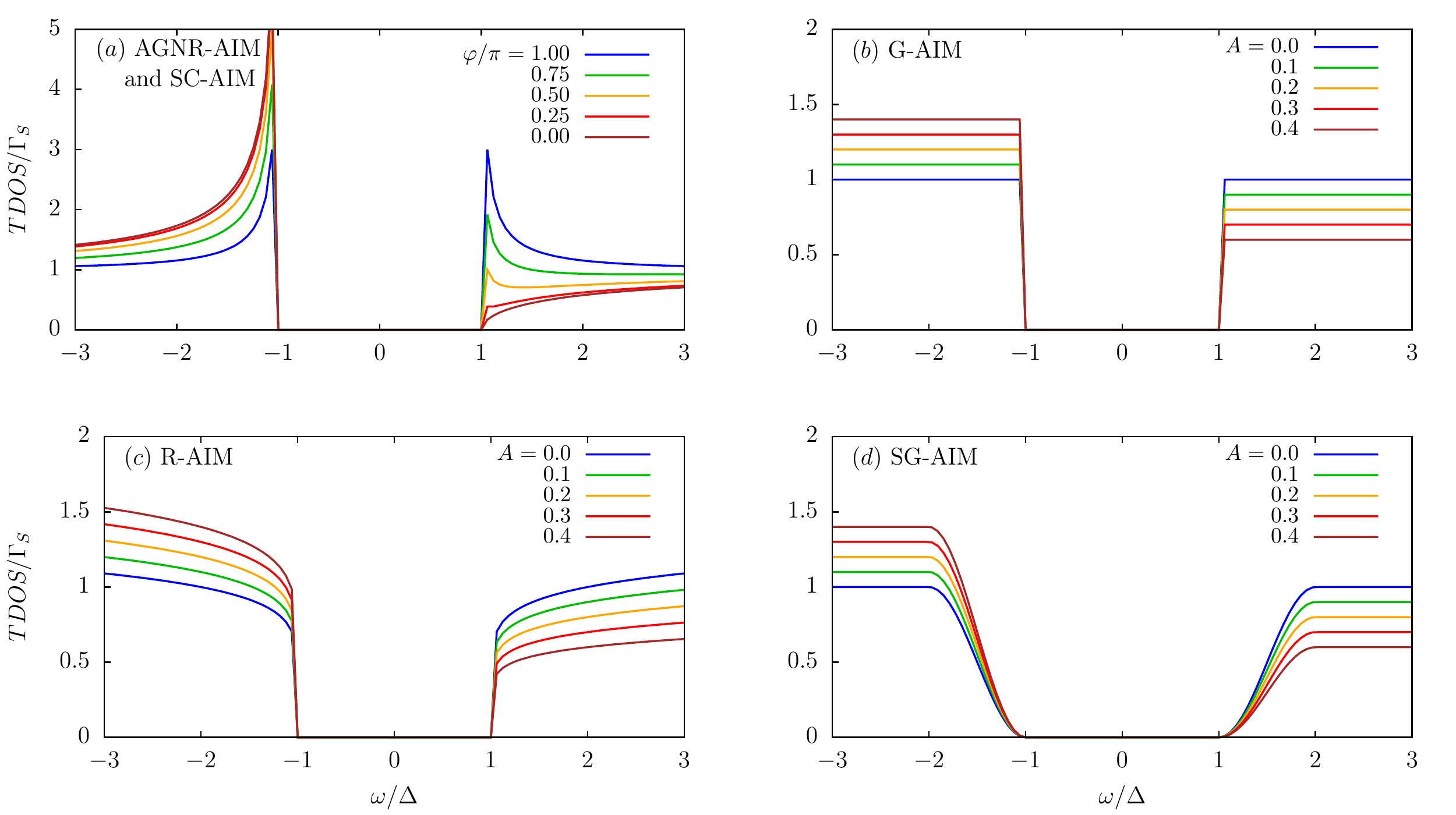}
	\caption{$(a)$-$(d)$ TDOS functions of four distinct semiconductor models with various properties. BCS-like divergences on the gap edges appear in panel $(a)$ and are typical for SC-AIM and AGNR-AIM cases given in Eq.~(\ref{BCS_DOS_w}). Gapped Anderson model (G-AIM) possesses only finite discontinuities on the gap edges as shown in panel $(b)$ and is given in Eq.~\eqref{DOS_flat}. Continuous TDOS functions appear when root functions are considered (R-AIM case) as given in Eq.~\eqref{DOS_root}) and are shown in panel $(c)$. Smoothed-out gap edges of the gapped Anderson model are considered in Eq.~\eqref{DOS_cos} and are shown in panel $(d)$. In SC-AIM and AGNR-AIM \textit{ph}-asymmetry is incorporated via parameter $\varphi$ that modulates the TDOS in the vicinity of the gap edges. In the remaining models \textit{ph}-asymmetry is incorporated via the parameter $A$. 
    \label{fig_0}}
\end{figure*}

Low temperature nanostructures, e.g., single molecules or various types of semiconducting quantum dots (QDs) coupled to a large system, represent a highly controllable testbed for various concepts of heavy fermion physics. Using host bands with metallic characteristics, one can then recreate and exert control over the Kondo effect~\cite{Kouwenhoven-2001,Agrait-2003,Wernsdorfer-2010,Rodero-2011,Meden-2019,Prada-2020,Evers-2020,Zonda-2021} known originally to appear in metals with diluted magnetic impurities~\cite{Hewson-1993}.
Its presence is marked by a zero frequency anomaly in the spectral function. The effective description is possible via the Anderson impurity model (AIM)~\cite{Anderson-1961} with its characteristic hierarchy of coupled energy scales. Thus, unbiased analytic solutions are only given via the Bethe ansatz~\cite{Andrei-1983}, while non-perturbative Renormalization Group (RG)~\cite{Metzner-2012,Kopietz-2010,Kopietz-2013} requires additional approximations. More general perturbative solutions~\cite{Hewson-RG-2013,Andergassen-2008,Janis-2007,Janis-2019} capture typically only qualitative aspects of the Kondo physics. Numerically, Quantum Monte Carlo (QMC) techniques provide the most flexible methods but are limited to finite temperatures and require ill-defined analytical continuation for spectral properties~\cite{Gull-2011, Gubernatis-2016}. In contrast, applying RG transformation numerically, i.e., via numerical renormalization group (NRG), offers unbiased access to wide parameter ranges~\cite{Wilson-1975,Krishna-1980a,Krishna-1980b,Bulla-Rev-2008}.Overall, a good agreement with the experimental data is observed~\cite{Goldhaber-Gordon-1998,Bulla-Rev-2008,pavarini2015many-body,Hewson-1993,Anderson-1961,Zonda-2021}.

Opening a spectral gap $\Delta$ in the host band, as routinely investigated in systems with superconducting leads, interferes with the energy cascade. Yet, NRG methods still offer predictions  \cite{Meden-2019,Meng-2009,Bauer-2007,Tanaka-2007,Tanaka-2007-jpsj,Zitko-2016,Zonda-2016,Kadlecova-2019} which are in excellent agreement with the experimental observations \cite{Chang-2013,Pillet-2010,Pillet-2013,Delagrange-2015,Delagrange-2016}. The missing conduction states in the gap region are known to change the parity of the ground state with the resulting $0$-$\pi$ quantum phase transitions (QPTs) being consequently observed~\cite{Yoshioka-2000,Tanaka-2007,Kadlecova-2017,Kadlecova-2019,Rodero-2011,Meden-2019,Rodero-2011,Meden-2019,Luitz-2012,Rozhkov-1999,Siano-2004,Novotny-Rossini-2005,Karrasch-2008}. The $0$-phase shows a singlet ground state, while a doublet ground state is observed in the $\pi$-phase. Moreover, the $0$-$\pi$ QPT is always accompanied by the crossing of the in-gap excited states known as Andreev bound states (ABSs).

On the other hand, gapped hosts without pairing correlations, e.g., narrow gap semiconductors, have attracted much smaller attention experimentally and are not fully understood theoretically either. For half-filled QD, there is a growing consensus that a particle-hole (\emph{ph}) symmetric gapped host band allows only a doublet ground state, while away from half-filling of the QD a critical $\Delta_c$ always exists below which the ground state is the Kondo singlet. These results have been established perturbatively \cite{Galpin-2008,Galpin-2008a}, as well as non-perturbatively by QMC \cite{Takegahara-1992} and NRG \cite{Moca-2010,Chen-1998,Pinto-2015}. Nevertheless, finite a spectral gap $\Delta$ makes the logarithmic discretization of the band, upon which NRG relies, to fail~\cite{Chen-1998,Moca-2010,Pinto-2015} and NRG iterations are then halted. Typically, additional, strictly speaking non-existent, NRG iterations are defined artificially to push the truncation temperature to $T<\Delta$.

Moreover, a very limited attention has been given to the relation between the QPTs in superconductor and semiconducting hosts. Of course, both systems exhibit gaps of radically different origin with the Josephson effect ensuring an exceptional flow of the current between superconducting leads even in equilibrium -- an effect that has no counterpart in semiconductors. The two systems could thus easily be deemed as completely distinct. However, in Ref.~\cite{Zalom-2021} a half-filled (HF) superconducting AIM (SC-AIM) has been exactly mapped via unitary transformation $\mathbb{T}$ onto a QD coupled to a scalar semiconducting band with a specific tunneling density of states (TDOS) as shown in Fig.~\ref{fig_0}$(a)$ which has the same shape as the TDOS of QD coupled to a semiconducting host band of armchair graphene nanoribbon (AGNR) \cite{Diniz-2020} with the resulting Hamiltonian referred to as AGNR-AIM here.

Since the reformulation of the SC-AIM problem according to Ref.~\cite{Zalom-2021} is performed via unitary transformation, the many-body spectra of both problems are exactly the same~\cite{Zalom-2021a}. The presence of \emph{ph}-asymmetry in the TDOS is then known to govern the QPT in SC-AIM as well as AGNR-AIM \cite{Zalom-2021,Zalom-2021a}. While in the SC-AIM case it is the  \emph{ph}-asymmetry of the TDOS obtained after the transformation $\mathbb{T}$ that leads to the $0$-$\pi$ transition, for AGNR-AIM, it is directly the electronic \emph{ph}-asymmetry~\cite{Zalom-2021a}. 

Nevertheless, it is unclear how robust such an equivalence is. We thus first, in-line with Refs.~\cite{Chen-1998,Moca-2010,Pinto-2015}, construct the gapped AIM (G-AIM) which has a finite step-like discontinuity on the gap edges as shown in Fig.~\ref{fig_0}$(b)$. Since, in the noninteracting case it is well known that for continuous TDOS a critical depth or width of the gap must be reached in order to have a bound state~\cite{economou2006green} we also construct two classes of continuous semiconductor TDOS functions for the interacting cases. The first is modeled via root behavior as shown in Fig.~\ref{fig_0}$(c)$ (R-AIM), while the other is obtained from G-AIM by smoothing out its discontinuity according to Fig.~\ref{fig_0}$(d)$ (SG-AIM). Since the \emph{ph}-asymmetry of the TDOS in the AGNR-AIM case was identified to trigger the QPT for half-filled QD, we insert the analogous behavior via the parameter $A$ into G-AIM, R-AIM and SG-AIM as shown in Fig.~\ref{fig_0} and discussed later in detail.

To circumvent the difficulties arising from truncation strategies of Refs.~\cite{Chen-1998,Moca-2010,Pinto-2015} we add metallic leads with constant broadening $\Gamma_M$ to the models and indicate these as SC-AIM+M, AGNR-AIM+M, G-AIM+M, S-AIM+M and R-AIM+M respectively. Standard NRG techniques may then be applied~\cite{Zalom-2021, Diniz-2020, Zalom-2021a}, but come on the expense of always inducing a singlet ground state to the system, which thus obscures the QPTs we want to observe at $\Gamma_M=0$. Nevertheless, by lowering $\Gamma_M$ to the smallest possible value guarantying numerical stability of NRG, we may still observe the in-gap states as slightly broadened subgap peaks as already shown for SC-AIM+M and AGNR-AIM+M \cite{Zalom-2021, Zalom-2021a}. Changing \emph{ph}-asymmetry via $A$ is then expected to cause shifts of the broadened in-gap peaks (if they are present) in analogy to the SC-AIM+M and AGNR-AIM+M. Their crossing of the Fermi energy signals, by analogy to the SC-AIM case, an underlying QPT present for $\Gamma_M=0$. Consequently, these QPTs can be detected indirectly.

The outline of this paper is as follows. We first review SC-AIM+M results in Sec.~\ref{subsec_bcs} and then define the four different semiconductor TDOS functions in Sec.~\ref{subsec_semi}. A description of the standard NRG procedure accompanied with a brief review of the basic notion of RG fixed points is given in Sec.~\ref{subsec_nrg}. Results on temperature-dependent effective behavior are first given in a detailed fashion for G-AIM with  \emph{ph}-symmetric TDOS in Sec.~\ref{subsec_reentrant} and then analogously for \emph{ph}-asymmetric choice in Sec.~\ref{subsec_reentrant_asym}. A brief comparison of all semiconductor cases is then undertaken using their reentrant thermodynamic behavior in Sec.~\ref{subsec_reentrant_compar}. To understand the nature of the underlying QPTs, we first analyze the subgap spectrum of G-AIM in Sec.~\ref{subsec_spec_AIM}. In Sec.~\ref{subsec_spec_compar} we then finally confirm that singlet-doublet transitions are supported by all here investigated semiconductor cases. However, as shown in Sec.~\ref{subsec_escape} for a sufficiently large depopulation of the regions adjacent to the gap edges, in-gap states leak out of the gap region and no final conclusion regarding the possibility of the underlying QPTs can be given. Finally, we argue in  Sec.~\ref{sec_exps} that a possible experimental realization of singlet-doublet transition with semiconductor host bands might require QD set out-of-half filling. The main conclusions are then summarized in Sec.~\ref{sec_conc}.

\begin{figure*}[ht]
	\includegraphics[width=2.00\columnwidth]{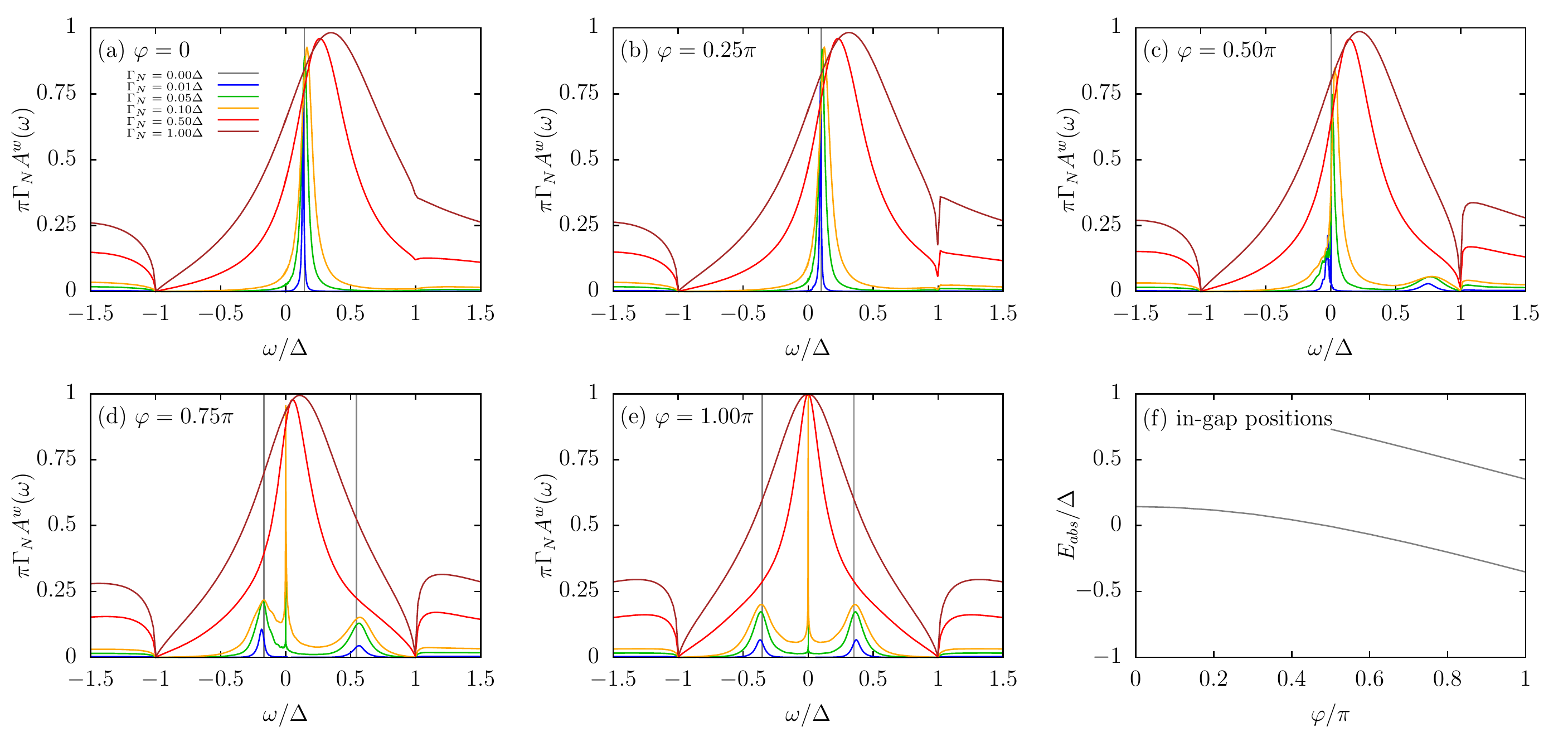}
	\caption{$(a)$-$(e)$ The in-gap part of the spectral functions of the generalized particle-hole asymmetric AGNR-AIM+M model shows broadened in-gap states for selected values of $\varphi$ as a parametric plot of $\Gamma_M$. In generalized AGNR-AIM+M model, $\varphi$ just parametrizes particle-hole asymmetry of the generalized AGNR band. Panels $(a)$-$(e)$ conform to the notation of Fig.~\ref{fig_2} and have been obtained at the same parameters. However, panel $(f)$ shows only the extracted position of the maxima of the in-gap peaks from panels $(a)$-$(e)$ as no dedicated NRG algorithm for gapped scalar host bands exists to date. \label{fig_1}}
\end{figure*}

\begin{figure*}[ht]
	\includegraphics[width=2.00\columnwidth]{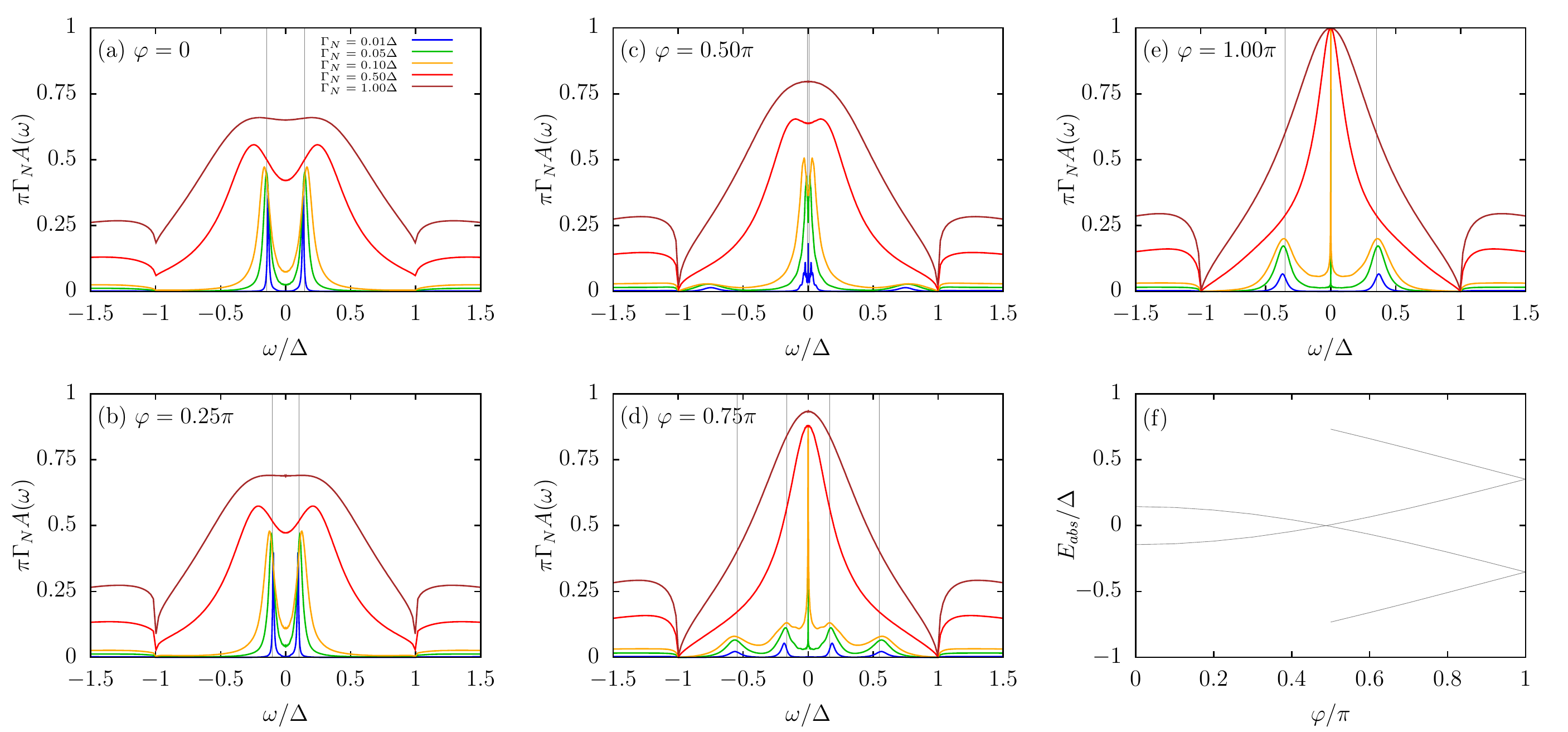}
	\caption{$(a)$-$(e)$ The in-gap part of the spectral functions shows broadened ABS states of the hybrid normal-superconducting reservoir for selected values of $\varphi$ as a parametric plot of $\Gamma_M$. The vertical gray lines show the positions of the sharp ABS states for SC-AIM of the corresponding model parameters. Clearly, taking the limit $\Gamma_M \rightarrow 0$ allows one to identify the position of sharp ABS states when the metallic electrode is completely decoupled from the problem. $(f)$ $\varphi$ dependence of sharp ABS states of SC-AIM (gray) and the position of maxima in their broadened ABS analogs of the hybrid normal-superconducting reservoir.  All panels have been calculated using NRG Ljubljana for $\Gamma_S=\Delta$, $\Delta=5 \cdot 10^{-4}B$ with $2B$ being the width of the band. For purely superconducting case $\Lambda=4$ while for the hybrid normal-superconducting reservoir $\Lambda=2$. \label{fig_2}}
\end{figure*}

\section{Theory \label{sec_theory}}

\subsection{Properties of SC-AIM+M model \label{subsec_bcs} }

Impurity models with superconducting host bands are successfully modeled using the SC-AIM~\cite{Wernsdorfer-2010,Rodero-2011,Luitz-2012,Meden-2019,Prada-2020} with the impurity treated as a spin $1/2$ Anderson orbital populated by the electrons $d_{\sigma}$ of spin projections $\sigma \in \{ \uparrow, \downarrow\}$. The corresponding Hamiltonian reads
\begin{eqnarray}
H_\mathrm{dot}
&=&
\sum_{\sigma} 
\varepsilon_{\mathrm{dot}}
d^{\dagger}_{\sigma}
d^{\vphantom{\dagger}}_{ \sigma}
+
U
d^{\dagger}_{\uparrow}
d^{\vphantom{\dagger}}_{ \uparrow}
d^{\dagger}_{\downarrow}
d^{\vphantom{\dagger}}_{ \downarrow},
\label{H_d}
\end{eqnarray}
where $d^{\dagger}_{\sigma}$ creates an electron of spin $\sigma$ on the QD while $d^{\vphantom{\dagger}}_{\sigma}$ annihilates it. The first term describes the energy level $\varepsilon_{\mathrm{dot}}$ and the second one the local Coulomb repulsion characterized by $U$. The half-filled point of the QD is therefore set by $\delta=0$ where $\delta=\varepsilon_{\mathrm{dot}}+U/2$.

The effect of the superconducting reservoir on the QD can be expressed in the following two equivalent ways. One may either formulate the corresponding BCS Hamiltonian for the band and the tunneling term, see for example Eqs.~(2) and (3) in Ref.~\cite{Zalom-2021}, or one may simply define a corresponding self-energy contribution $\Sigma$. Its imaginary part, the TDOS function $\Gamma(\omega)$, carries all the necessary information as the real part of $\Sigma$ follows from Kramers-Kronig relations. In this paper, we will work in the second formalism.

In the SC-AIM+M model, we assume a constant hybridization $\Gamma_M$ of the metallic lead to the QD and right and left BCS electrodes with equal hybridizations $\Gamma_L$ and $\Gamma_R$ which according to Ref.~\cite{Kadlecova-2017} can be transformed by analytic means into any asymmetric configuration of $\Gamma_L$ and $\Gamma_R$. Using the Nambu spinor formalism with
\begin{equation}
D^{\dagger} 
=
\left(
d^{\dagger}_{\uparrow},
d^{\vphantom{\dagger}}_{\downarrow}
\right),
\end{equation}
the tunneling self-energy becomes
\begin{equation}\label{eq:self-energy}
\mathbb{\Sigma}^D(\omega^+)
=
\Sigma^D_n(\omega^+) \mathbb{1} 
+ 
\Sigma_a^D (\omega^+) \mathbb{\sigma}_x.
\end{equation}
Here, $\mathbb{1}$ is a two by two unit matrix, $\mathbb{\sigma}_x$ is the Pauli matrix $x$ and the self-energy components read
\begin{eqnarray}
\Sigma^D_n(\omega^+)
&=&
-i\Gamma_M
+ \,
\Gamma_S \omega F(\omega^+),
\\
\Sigma^D_a(\omega^+)
&=&
\Gamma_S\Delta\cos\left(\frac{\varphi}{2}\right)
F(\omega^+)
\end{eqnarray}
with
\begin{equation}
F(\omega^+)
=
\begin{cases}
-\frac{2}{\pi\sqrt{\Delta^2-\omega^2}} \arctan
\left(\frac{B}{\sqrt{\Delta^2-\omega^2}}\right),
&
\text{for }|\omega|<\Delta
\\
-\frac{i\, \mathrm{sgn}(\omega)}{\sqrt{\omega^2-\Delta^2}}
+
\frac{\ln \left( \frac{B+\sqrt{\omega^2-\Delta^2}}{B-\sqrt{\omega^2-\Delta^2}}\right)}{\pi\sqrt{\omega^2-\Delta^2}},
&
\text{for }\Delta<|\omega|<B
\end{cases}
\label{eq_6}
\end{equation}
where $\Delta$ is the BCS gap, $\varphi$ is the phase drop across the BCS leads and $B$ is the half-width of the bands. In NRG we set $B$ much larger that any other energy scale, typically $B>10^3\Delta$.  

While Eqs.~\eqref{H_d}-\eqref{eq_6} constitute a complete theory for the QD (by setting $\Gamma_M=0$ also for SC-AIM), they cannot be solved analytically once $U \neq 0$. A direct application of the NRG algorithm with logarithmic discretization to the general superconducting case with $\varphi \neq 0$ is only possible for SC-AIM with $\Gamma_M=0$ \cite{Satori-1992,Yoshioka-2000} while for $\Gamma_M \neq 0$ a total number of three different electronic bands require the interleaved discretization \cite{Mitchell-2014} which is however not providing sufficient precision. Following Ref.~\cite{Zalom-2021} we reformulate the theory given in Eqs.~\eqref{H_d}-\eqref{eq_6} by rotating the $d_{\sigma}$ electrons of the QD via a unitary transformation $\mathbb{T}$ as
\begin{eqnarray}
w^{\dagger}_{\uparrow}
&=&
\sqrt{\frac{1}{2}}
\left(
-d^{\dagger}_{\uparrow}
+
d^{\dagger}_{\downarrow}
\right),
\\
w^{\dagger}_{\downarrow}
&=&
\sqrt{\frac{1}{2}}
\left(
d^{\dagger}_{\uparrow}
+
d^{\dagger}_{\downarrow}
\right).
\end{eqnarray}
The resulting Bogolyubov-like quasi-particles $w_{\sigma}$ set a new reference basis in which $H_{\mathrm{dot}}$ reads
\begin{equation}
H_{\mathrm{dot}}
=
\sum_{\sigma} 
\varepsilon
w^{\dagger}_{\sigma}
w^{\vphantom{\dagger}}_{ \sigma}
+
U
w^{\dagger}_{\uparrow}
w^{\vphantom{\dagger}}_{ \uparrow}
w^{\dagger}_{\downarrow}
w^{\vphantom{\dagger}}_{ \downarrow}
+
\Delta_{\mathrm{dot}}
\left(
w^{\dagger}_{\downarrow}
w^{\dagger}_{\uparrow}
+
w^{\vphantom{\dagger}}_{ \uparrow}
w^{\vphantom{\dagger}}_{ \downarrow}
\right)
\label{H_w}
\end{equation}
with $\Delta_{\mathrm{dot}} =
\left(
\varepsilon_{\mathrm{dot}}+U/2
\right)$ and the corresponding term being thus understood as induced superconductivity present only for QDs out-of-half-filling. At half-filling $\Delta_{\mathrm{dot}}=0$ and the transformation $\mathbb{T}$ leaves the $H_\mathrm{dot}$ form-invariant.

Regardless of the filling of the QD, the TDOS function becomes diagonal and symmetric with respect to $w_{\uparrow} \leftrightarrow w_{\downarrow}$ (see Fig.~\ref{fig_0}):
\begin{equation}
\Gamma^w_\textrm{BCS}(\omega;\varphi)
=
\Gamma_M
+
\frac{\Gamma_S|\omega|\Theta(\omega^2-\Delta^2)}
{\sqrt{\omega^2-\Delta^2}}
\left
(1-\frac{\Delta }{\omega}\cos\frac{\varphi}{2}
\right).
\label{BCS_DOS_w}
\end{equation}
The TDOS function has typical BCS-like singularities at the gap edges $\omega=\pm \Delta$ and is \emph{ph}-asymmetric with respect to the Bogolyubons $w$ for $\varphi \neq \pi$. We also stress that although \emph{ph}-asymmetry affects essentially only the proximity of the gap region, it was shown in Ref.~\cite{Zalom-2021} to be the underlying cause of the $0$-like and $\pi$-like regimes present in SC-AIM+M. However, out-of-half-filling, the presence of the $\Delta_{\mathrm{dot}}$-term lowers the symmetry of the initial and subsequent NRG iterations and leads to a more elaborated phase space which is avoided in this paper by demanding half-filled QD.

The \emph{ph}-asymmetry of the TDOS causes the resulting spectral functions $A^w$ to be also \emph{ph}-asymmetric for $\varphi \neq \pi$ as shown in Fig.~\ref{fig_1}. However, for SC-AIM+M such spectral functions are unaccessible as only the tunneling of electrons can be measured in STM experiments. It is thus necessary to express all results using the $d_{\sigma}$ electrons as easily performed via Eq.~(34) of Ref.~\cite{Zalom-2021}. The normal spectral functions $A^d_n$ of $d_{\sigma}$ electrons are obtained by symmetrizing its counterpart $A^w_n$ from the $w_{\sigma}$ basis. As shown in Fig.~\ref{fig_2}, as long as $\Gamma_M \lesssim \Delta/5$ (weakly coupled metallic electrode), the effect of metallic electrode on the subgap portion of $A^d_n$ is limited to a mere broadening of the otherwise sharp in-gap peaks of SC-AIM and the appearance of Kondo peak for $\varphi>\varphi^*$ where they cross with Fermi energy occurs, see panel \ref{fig_2}$(f)$. Therefore, although the presence of weakly coupled metallic electrode inevitably forces singlet ground state into the system at low $T$ we may still deduce the nature of QPTs for SC-AIM indirectly by observing the crossing of the broadened ABS states of SC-AIM+M with Fermi energy $\Gamma_M\neq 0$~\cite{Zalom-2021,Zalom-2021a}.

\subsection{Semiconducting reservoirs augmented by a normal metallic lead \label{subsec_semi}}

In general, \emph{ab initio} methods reliably deliver densities of states in the bulk of realistic semiconductor materials~\cite{xiao2011acuurate,crowley2016resolution}. 
However, its interaction with a given nanosystem of interest gives in principle rise to realization-dependent TDOS functions~\cite{Appelt-2018}. Therefore, instead of investigating specific TDOS functions resembling the bulk properties, we construct some prototypical TDOS cases with focus given to the general features essential for the existence of singlet-doublet QPTs. Realistic configurations are discussed briefly in Sec.~\ref{sec_exps}.

We start with the case assessed in Ref.~\cite{Diniz-2020}, where spin $1/2$ QD was coupled to a hybrid reservoir of metallic-semiconducting character. The proposed experimental realization included AGNR as the substrate while a weakly coupled metallic electrode could stem from the corresponding STM tip. Besides the experimental motivation, it also realizes a case with infinite TDOS discontinuity on the gap edge. The TDOS of such AGNR-AIM+M case was previously shown in Fig.~\ref{fig_0}(a) and is prescribed as
\begin{equation}
\Gamma_\textrm{AGNR+M}(\omega)
=
\Gamma_M
+
\frac{\Gamma_S|\omega|\Theta(\omega^2-\Delta^2)}
{\sqrt{\omega^2-\Delta^2}},
\label{DOS_AGNR}
\end{equation}
where $\Delta$ is the semiconductor gap of AGNR, $\Gamma_S$ ($\Gamma_M$) is the  hybridization to the QD to the semiconductor (metallic) host. $H_{\mathrm{dot}}$ is given by \eqref{H_w} without the $\Delta_{\mathrm{dot}}$ which relates only to the SC-AIM or SC-AIM+M case. This TDOS is, however, just a specific case of the TDOS characterizing the normal-superconducting reservoir for $\varphi=\pi$ written in its scalar basis $w$ as pointed out in Ref.~\cite{Zalom-2021a}. Following Ref.~\cite{Zalom-2021a}, we can generalize the AGNR-AIM+M case by including $\varphi$-dependence which models the \emph{ph}-asymmetry of the band to exactly match the form of \eqref{BCS_DOS_w}: $\Gamma_{\textrm{AGNR-AIM+M}}(\omega;\varphi)\equiv\Gamma^w_\mathrm{BCS}(\omega;\varphi)$. In contrast to SC-AIM+M case, here $\varphi$ is just a measure of  \emph{ph}-asymmetry and as such does not refer to the phase of the leads.
Note that the overall number of states in $\Gamma_{\textrm{AGNR-AIM+M}}(\omega;\varphi)$ is equal to $2\Gamma_M B + 2 \sqrt{B^2-\Delta^2} \, \Gamma_S$ or $\approx 2\left(\Gamma_M+\Gamma_S \right)B$ in the wide band limit. Standard NRG has already been used on this system in the scope of solving the SC-AIM+M case. The corresponding spectral functions of AGNR-AIM+M are thus given by Fig.~\ref{fig_1}. The position of the resulting in-gap states matches the broadened ABSs of SC-AIM+M case. Due to the missing symmetrization [Eq.~(34) in  Ref.~\cite{Zalom-2021}], the in-gap peaks do not mix particle and hole properties together and their total number is always half of those found in SC-AIM+M. 

To test if BCS-like divergences on the gap edges are crucial for the development of the $0$-$\pi$ transition, we first consider a gapped AIM where the discontinuity is finite~\cite{Chen-1998, Moca-2010}. The \emph{ph}-asymmetry of the bands is explicitly built in via the parameter A as
\begin{equation}
\Gamma_\textrm{G-AIM+M}(\omega)
=
\begin{cases}
\Gamma_M
&
\text{for }|\omega|<\Delta,
\\
\Gamma_M + \left( 1-\mathrm{sgn}[\omega]A \right) \Gamma_S
&
\text{for }|\omega|\geq\Delta,
\end{cases}
\label{DOS_flat}
\end{equation}
with $0\leq A<1$ to ensure positive TDOS with the excess of states present in the electronic part (negative frequencies) for any $A \neq 0$. Obviously, $A=0$ represents a symmetric choice which has been already investigated in Refs.~\cite{Chen-1998, Moca-2010}. Note that the integral over all frequencies of the TDOS is fixed to $2\Gamma_M + 2 \left( B -\Delta \right)  \, \Gamma_S$ for arbitrary $A$. In the wide band limit ($\Delta \ll B$), the integral can be approximated by $2\left( \Gamma_M+\,\Gamma_S\right)B$ as for the AGNR-AIM+M case. 

We further soften the gap discontinuities and consider
\begin{equation}
\Gamma_\textrm{SG-AIM+M}(\omega)
=
\begin{cases}
\Gamma_M
& 
\hspace{-2.4cm} \text{for }|\omega|<\Delta,
\\
\Gamma_M + \left( 1 -\mathrm{sgn}[\omega]A\right)
\Gamma_S \times
& \\
\left[
1-\cos\left(\frac{\pi(|\omega|-\Delta)}{2(b-1)\Delta} \right)
\right]
& \\
\hspace{2.5cm}  \text{for }b\Delta\geq|\omega|\geq \Delta,
\\
\Gamma_M + \left( 1 -\mathrm{sgn}[\omega]A\right)
\Gamma_S
& \\
\hspace{3.3cm}  \text{for } |\omega|> b\Delta,
\end{cases}
\label{DOS_cos}
\end{equation}
where $b>1$ defines the width over which an otherwise present discontinuity of \eqref{DOS_flat} is smoothed out. Parameter $A$ governs the particle-hole asymmetry of the TDOS and is defined as previously. Due to the smoothening, some states are removed in the region $\Delta<|\omega|\leq b\Delta$ but in the wide band limit with moderate values of $b$ the integral over all frequencies of the TDOS is once again approximately $2\left( \Gamma_M+\,\Gamma_S\right)B$. We stress that strictly speaking, only the region $|\omega|<\Delta$ exhibits a full gap, however, the depletion of the TDOS for $b\Delta\geq|\omega|\geq \Delta$ effectively makes the gap region larger.

The last class of semiconductor TDOS functions is constructed to depopulate not only the proximity of the gap region but even more extended regions. We define it as
\begin{equation}
\Gamma_{\textrm{R-AIM}}(\omega)
=
\begin{cases}
\Gamma_M
& 
\hspace{-2.2cm}  \textit{for $|\omega|<|\Delta|$},
\\
\Gamma_M + \left( 1-\mathrm{sgn}[\omega]A \right)
\Gamma_S 
\left(
|\omega|-\Delta
\right)^{1/r}
& \\
\hspace{4.2cm}  \textit{for $\omega>\Delta$}.
\end{cases}
\label{DOS_root}
\end{equation}
Here, we demand $r>1$ with typical value set to $r=8$. Nevertheless, as $r$ governs the smoothness of the transition into the gap region, we show in Sec.~\ref{subsec_escape} that small values of $r$ may release in-gap states into the supragap region, which sets limits to a possible equivalence of QPTs in semiconductors to those of SC-AIM.

TDOS functions \eqref{DOS_root} with $A=0$ are similar to the bulk DOS of granular metallic systems at the onset of Coulomb blockade regime in two and three dimensions  \cite{Beloborodov-2004,Sun-2010}. Nevertheless, compared to other TDOS functions, the integral over all frequencies is in the wide band limit given as $2\Gamma_M B + 2 r \Gamma_S/(1+r)$. Therefore, to make quantitative comparisons with other TDOS, we use a scaled $\tilde{\Gamma}_S=(1+r)\Gamma_S/r$ to match the integral properties.

\subsection{NRG solution \label{subsec_nrg} }

All models considered here are tractable within the standard one-channel NRG in the energy representation~\cite{Bulla-1994, Bulla-Rev-2008}. The structured TDOS functions at $\Gamma_M \neq 0$ provide all necessary coefficients to construct the Wilson chains of the corresponding NRG calculations~\cite{Bulla-1994, Bulla-Rev-2008}. 
For SC-AIM ($\Gamma_M = 0$), reliable (converged) conventional NRG solutions are available for subgap properties \cite{Satori-1992,Yoshioka-2000}, however, even here artificial truncations are required to reconstruct the supragap spectral functions \cite{Zitko-2016}. For the $\Gamma_M \neq 0$ cases, NRG Ljubljana code~\cite{NRGzenodo} with the intertwined $z$-discretization according to the scheme introduced by \v{Z}itko and Pruschke~\cite{ZitkoPruschke-2009} was thus employed with the discretization parameter $\Lambda=2$ and $z= n/10$ with $n \in \{ 1, \ldots 10 \}$ with subsequent application of the so-called self-energy trick to the spectral functions~\cite{Bulla-1998}. 
\begin{figure}[ht]
	\includegraphics[width=1.00\columnwidth]{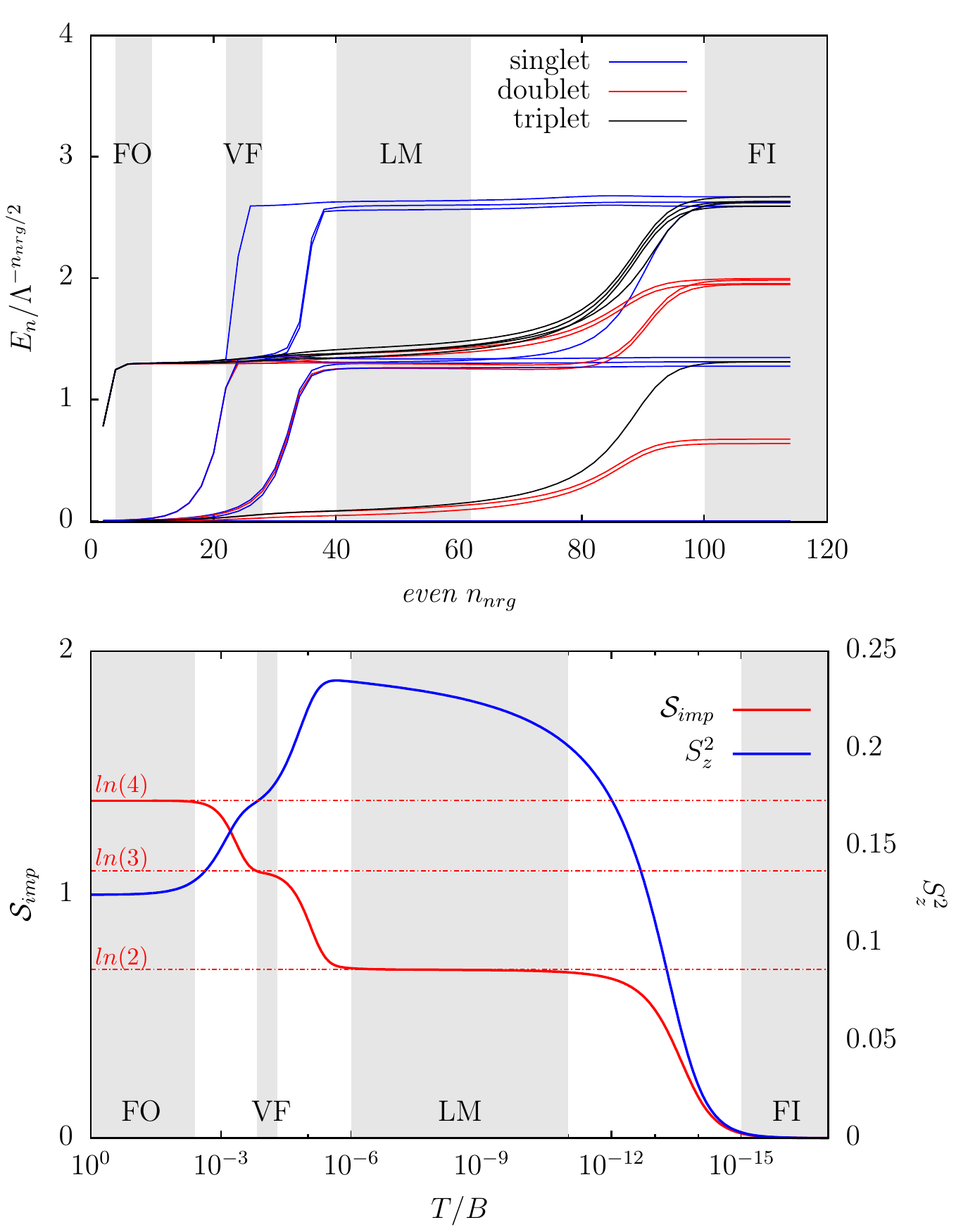}
	\caption{
	$(a)$
	Flow of rescaled eigenenergies $E_n$ during the even NRG iterations $n_{\textit{even}}$ of the asymmetric AIM with $U/B=10^{-3}$, $\Gamma_M/B=\pi/2 \cdot 10^{-6}$ and $\varepsilon_\mathrm{dot}/B=-2 \times 10^{-4}$. Discretization factor was set to $\Lambda=2$. Free orbital (FO) fixed point for $n_{nrg} < 10$, valence-fluctuation (VF) fixed point for $20<n_{nrg}<26$, local-moment (LM) fixed point for $40<n_{nrg}<64$ and frozen-impurity (FI) fixed point for $n_{nrg}>100$.
	$(b)$
	Temperature dependence of the impurity entropy $\mathcal{S}_{imp}$ with the contribution of the host band subtracted from the overall entropy of the system (red line) and $S_z^2$ expectation value (blue line). In the FO regime the impurity behaves as four fluctuating levels, i.e., $\mathcal{S}_{imp}=ln(4)$ while in the VF regime the doubly occupied state is effectively projected out and the impurity behaves as three level one resulting in $\mathcal{S}_{imp}=ln(3)$. Proceeding further in the flow, zero occupied states are effectively projected out leaving only two levels left, i.e., $\mathcal{S}_{imp}=ln(2)$. Finally, at very low $T$ FI fixed point means that the impurity is effectively frozen leaving thus $\mathcal{S}_{imp}=0$.
	\label{fig_4}}
\end{figure}

While dynamic properties, like spectral functions, might nowadays be obtained with great accuracy \cite{Bulla-Rev-2008} the original NRG idea of obtaining temperature-dependent effective description from the underlying microscopic model, as presented in seminal papers \cite{Krishna-1980a,Krishna-1980b}, is to a great extent marginalized. Yet, all thermodynamic and spectral properties are relatively straightforward consequences of specific subspaces of the QD being effectively projected out in the given fixed points of NRG transformations. These can be directly read-off from the corresponding flow diagrams. Because the effective behavior constitutes a major unifying feature for the gapped TDOS functions studied here, we briefly review this topic. Furthere details, including the definitions of the effective Hamiltonians and calculation of linear corrections to these, can be found in Refs.~\cite{Krishna-1980a,Krishna-1980b}.

We would like to stress that although we keep half-filling of the QD and vary only the  \emph{ph}-asymmetry of the host bands, the presence of the gap in the semiconductor host bands actually forces a previously not understood effective behavior which is in a close analogy to the out-of-half filling AIM with flat TDOS --- a case already treated by K. G. Wilson et al.~\cite{Krishna-1980b}. We show its RG flow of the energy eigenstates $E_n$ for even NRG iterations in Fig.~\ref{fig_4}(a). In accordance with the Ref.~\cite{Krishna-1980b} the RG flow approaches four fixed points of AIM with QD out-of-half-filling. In the order of decreasing $T$, they appear as the free-orbital (FO), valence-fluctuation (VF), local-moment (LM) and finally the frozen-impurity (FI) fixed point. When the QD is at half-filling the last fixed point is typically termed strong coupling (SC) fixed point. They all can be read off from the flow by noticing stabilized (steady) values of low lying $E_n$ and then comparing the relative positions of states with a given spin parity to each other~\footnote{Note, that in Fig.~\ref{fig_4}(a) we present the flow for the even NRG iterations, which have the advantage of correctly ascribing the quantum numbers to the ground state. Odd NRG iterations carry the same information about the effective behavior but the $E_n$ values, their quantum numbers and multiplicities completely differ compared to the picture with even iterations. In the end, the choice is arbitrary and just a matter of convenience.}.

The numerical positions of $E_n$ in FO, VF, LM and SC regimes depend only on $\Lambda$ and are thus independent of the model parameters as explained in Refs.~\cite{Krishna-1980a,Krishna-1980b}. The FI fixed point is the only regime, in which model parameters play a role. Nevertheless, for $\delta \approx 0$ it may be understood using the rescaled $E_n$ values of SC regime at $\delta=0$. Such a case is shown in Fig.~\ref{fig_4} and is marked by splitting of rescaled $E_n$ around the positions they assume at $\delta=0$. Taking however $\delta \rightarrow \infty$ the splitting becomes extremely strong. The lowest doublet state is then pushed against the ground state and the separation of the singlet ground state from the next doublet $E_n$ becomes extremely small. Note also, that the VF and LM fixed points can be skipped at certain values of $\varepsilon_\mathrm{dot}$ but the FI or SC fixed points are always present for $\delta>0$ and $\delta=0$, respectively. It is important to stress that systems close to RG fixed points show characteristic universal with corresponding effective Hamiltonians, as shown in Refs.~\cite{Krishna-1980a,Krishna-1980b}, offering then a simple yet precise characterization. 

Let us now briefly review the effective behavior in the FO, VF, LM, FI and SC regimes. The FO fixed point belongs to the most trivial ones because the system behaves as a free Anderson impurity that is disconnected from the host band. It can be obtained by setting $\Gamma=U=\varepsilon_\mathrm{dot}=0$. The VF fixed point that might follow after the FO fixed point appears only at certain values of $\varepsilon_\mathrm{dot}$ and corresponds to the situation where doubly occupied states on the QD are effectively projected out and as such corresponds to $\Gamma=\varepsilon_\mathrm{dot}=0$ and $U \rightarrow \infty$. The LM fixed point connects AIM to the Kondo model where the doubly and zero occupied states of the QD are projected out. The antiferromagnetic coupling $J$ of the effective Kondo model can be read out from the corresponding NRG calculations. In the final stages of NRG flow in the half-filled case we always observe the SC fixed point while out-of-half-filling FI emerges. In FI, all states except of the unoccupied QD are projected out. In the SC fixed point the QD freezes down to a single state and this state is a specific linear combination of singly occupied QD states. The resulting effective model resembles then the free-electron Hamiltonian. The impurity in the different regimes behaves thus effectively as an ensemble of $4$ (FO), $3$ (VF), $2$ (LM) and $1$ (FI,SC) fluctuating states with the resulting entropy values $ln(4)$, $ln(3)$, $ln(2)$ and $0$ respectively. In between the various regimes smooth crossovers are observed as shown in Fig.~\ref{fig_4}$(b)$ (red line). Fluctuations of the $S_z$ projection of spin onto the $z$ axis follow correspondingly as shown in Fig.~\ref{fig_4}$(b)$ (blue line).

\begin{figure*}[ht]
	\includegraphics[width=2.00\columnwidth]{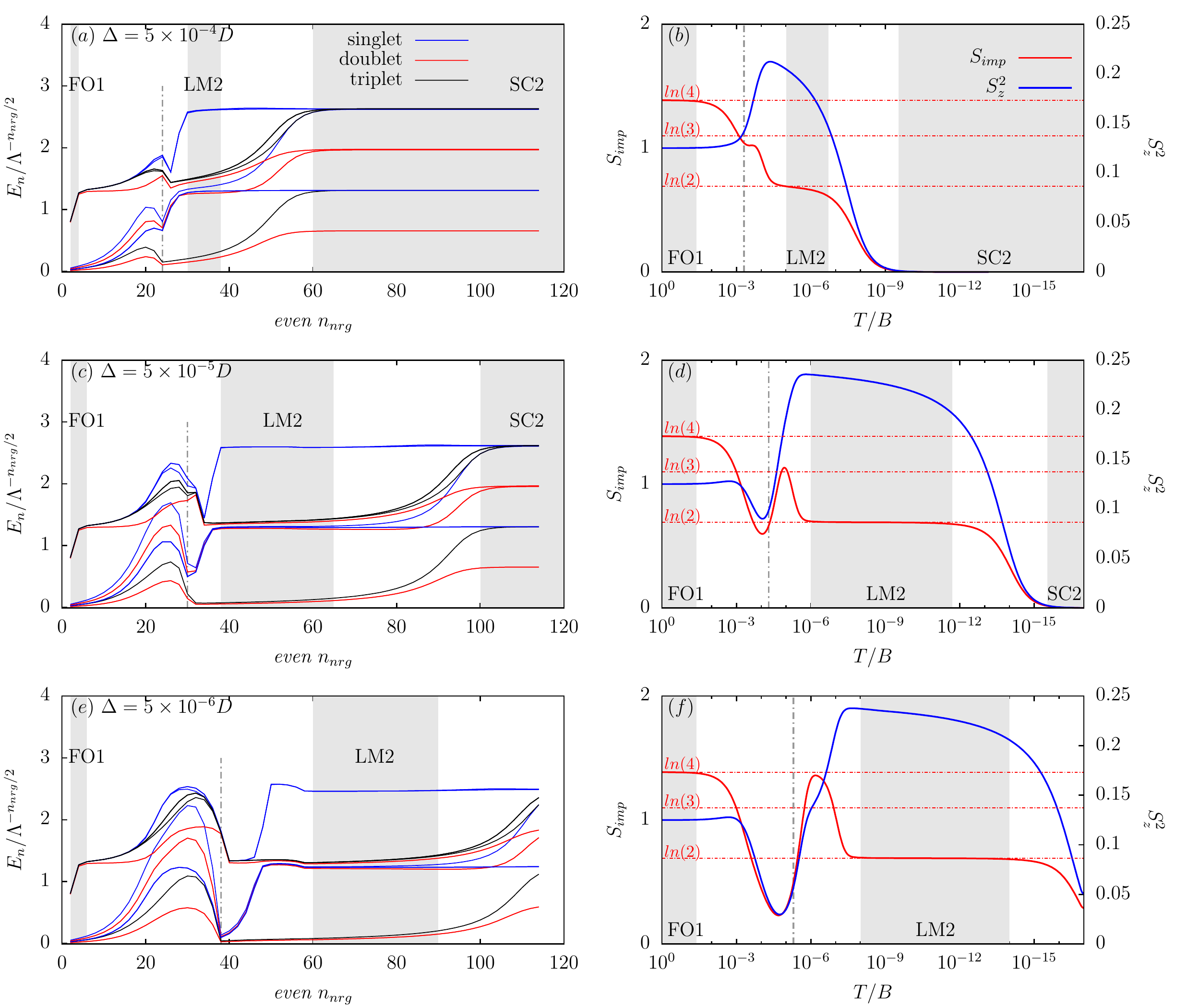}
	\caption{
	$(a)$
	Flow of rescaled $E_n$ during the iterative NRG calculation of model \eqref{DOS_flat} with $U/\Delta=3$, $\Gamma_S/\Delta=1$, $\Gamma_M \Delta =0.1$, $\Delta=5 \times 10^{-4}B$ and $A=0$. Discretization factor was set to $\Lambda=2$.
	$(b)$
	Temperature dependence of the impurity entropy $S_\mathrm{imp}$ (contribution of the host band subtracted from the overall entropy of the system) and $S_z^2$ for the same parameters as in panel $(a)$.
	$(c)$
	Flow of rescaled eigenenergies $E_n$ during the iterative NRG calculation of model \eqref{DOS_flat} with parameters as in panel $(a)$, only $\Delta=5 \times 10^{-5}B$
	$(d)$
	Temperature dependence of the impurity entropy $S_\mathrm{imp}$ and $S_z^2$ for the same parameters as in panel $(c)$.
	$(e)$
	Flow of rescaled eigenenergies $E_n$ during the iterative NRG calculation of model \eqref{DOS_flat} with parameters as in panel $(a)$, only $\Delta=5 \times 10^{-6}B$
	$(f)$
	Temperature dependence of the impurity entropy $S_\mathrm{imp}$ and $S_z^2$ for the same parameters as in panel $(e)$.
	Reentrant nature of the flow is apparently visible, flow of $E_n$ at high $T$ is initially in the free orbital regime (FO1), only in panel $(e)$ traces of strongly coupled high $T$ regime (SC1) are visible. Dash-dotted line separates the beginning of the low $T$ flow which might experience the low $T$ free orbital (FO2) regime  as in panel $(e)$ that continues into the local-moment (LM2) regime and finally flows into the strongly-coupled regime (SC2) for all $\Delta$ values. The thermodynamic properties have the corresponding behavior of the fixed point regimes.
	\label{fig_5}
	}
\end{figure*}

\section{Temperature-dependent effective models and thermodynamic properties \label{sec_thermo} }

In this section, we first examine the RG flow of rescaled eigenenergies $E_n$ for G-AIM with \emph{ph}-symmetric TDOS at $A=0$ and demonstrate the corresponding effective behavior of the system using selected thermodynamic properties --- namely the entropy of the impurity $\mathcal{S}_\mathrm{imp}$ and the mean value of spin projection operator $S_z^2$. We then repeat the same analysis for G-AIM with $A=0.1$ and show its remarkable correspondence to AIM with QD out-of-half-filling. We stress that the rest of the TDOS cases as well as the superconducting realization from Sec.~\ref{subsec_bcs} follow the same behavior as shown by comparison of the thermodynamic properties in Sec.~\ref{subsec_reentrant_compar}. 

\subsection{Reentrant behavior of  \emph{ph}-symmetry G-AIM \label{subsec_reentrant}}
 
In Fig.~\ref{fig_5} we present the RG flow of rescaled eigenenergies $E_n$ for G-AIM with $A=0$ (\emph{ph}-symmetric TDOS) at three selected values of $\Delta$. In all cases, the flow consists of two sectors which connect approximately at $n_\Delta=-2\ln\Delta/\ln\Lambda$ where the energy resolution of NRG iterations reaches approximately $\Delta/B$. In Fig.~\ref{fig_5}, this is marked by vertical dash-dotted lines. For NRG iterations $n_\mathrm{nrg}<n_{\Delta}$ we observe the high $T$ portion of the flow that abruptly changes the character at $n_\mathrm{nrg}=n_{\Delta}$ where it transits into the low temperature behavior observed for $n_\mathrm{nrg}>n_{\Delta}$.

The high temperature portion of the RG flow is in one to one correspondence with that of the ordinary AIM with flat TDOS and QD at half filling. Let us note that in the high temperature part of the flow only the FO effective regime is clearly present (marked as FO1 in Fig.~\ref{fig_5}) for all selected values of $\Delta$, while only at $\Delta =5 \times 10^{-6}B$ the first traces of the high temperature strongly coupled regime (SC1) show up (see for example the local minimum at $T/B\approx 10^{-5}$ in  Fig.~\ref{fig_5}(f)). Due to the selected values of  $U/\Delta=3$ and $\Gamma_S/\Delta=1$ the LM fixed points are completely skipped. The numerical calculations for even smaller $\Delta$ as presented here are challenging. Nevertheless, from the evolution of $E_n$ flow with decreasing $\Delta$ one can extrapolate that at very small values of $\Delta$ the system can reach the high-$T$ strongly coupled regime SC1. This is in compliance with predictions based on the results presented in Ref.~\cite{Diniz-2020} for the AGNR-AIM+M case and in Ref.~\cite{Zalom-2021a} for the SC-AIM+M case. 

\begin{figure*}[ht]
	\includegraphics[width=2.00\columnwidth]{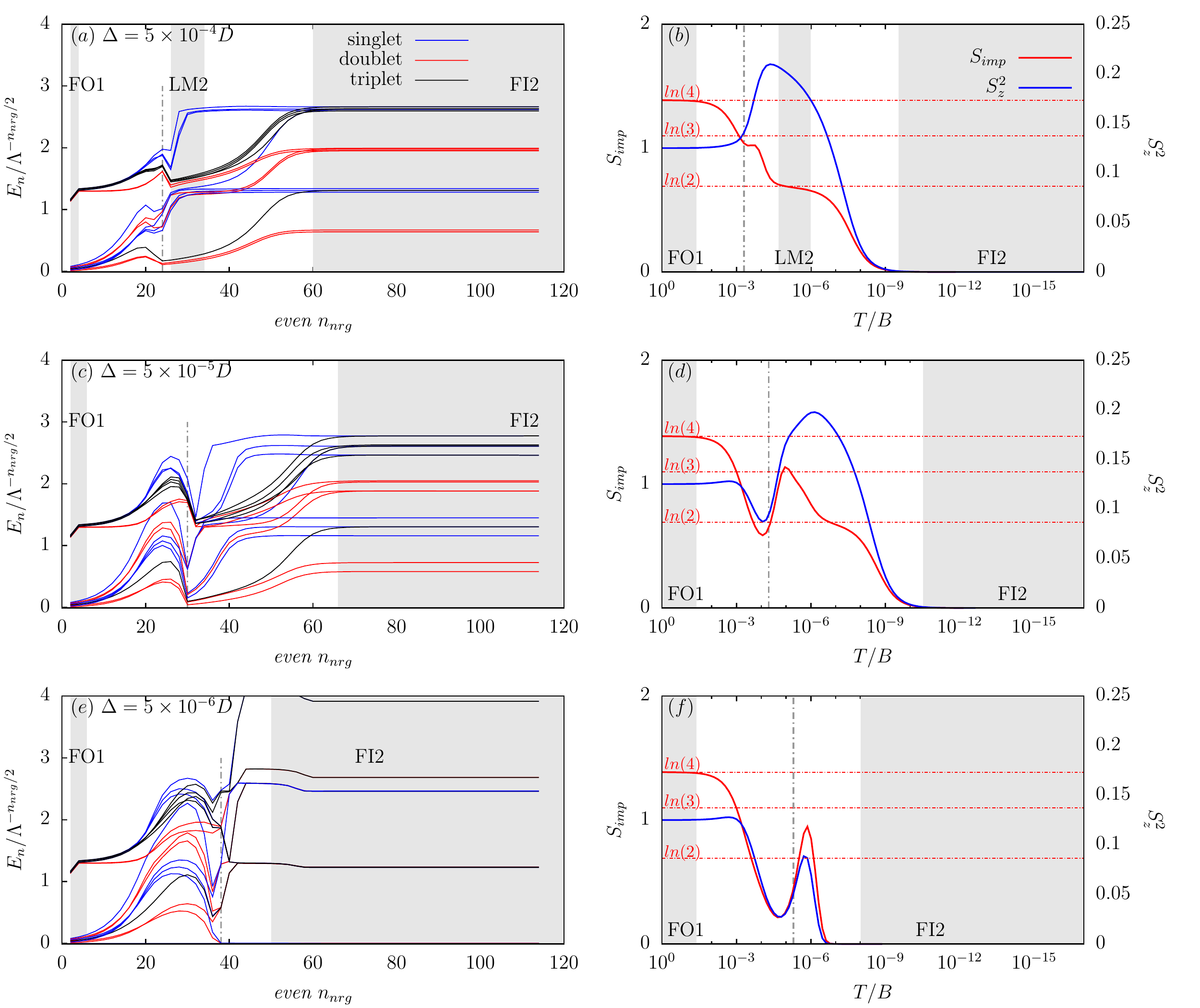}
	\caption{
		Panels $(a)$-$(f)$ correspond to their counterparts in Fig.~\ref{fig_5} with only the \emph{ph}-asymmetry factor set to $A=0.1$.
		Reentrant nature of the flow is preserved also for G-AIM with  \emph{ph}-asymmetry but the resulting effective regimes correspond to AIM with flat TDOS and QD out-of-half-filling. Thus, the flow of rescaled $E_n$ at high $T$ is initially in the free orbital regime (FO1), only in panel $(e)$ traces of frozen-impurity high $T$ regime (FI1) are visible. Dash-dotted line separates the beginning of the low $T$ flow which might experience the local-moment (LM2) regime and finally flows into the frozen-impurity regime (FI2) for all $\Delta$ values. The thermodynamic properties correspond to their effective models as discussed in Fig.~\ref{fig_4}.
		\label{fig_6}
	}
\end{figure*}

However, regardless of the existence of high temperature strongly coupled regime, we observe at $n_{nrg}\approx n_{\Delta}$ an abrupt change into a different effective mode which is caused by the reentrancy known also from previous studies of the AGNR-AIM+M~\cite{Diniz-2020} and SC-AIM+M~\cite{Zalom-2021a}. At very small $\Delta$ values the low temperature portion of RG flow initially develops traces of the free orbital regime (FO2). At values $\Delta \approx 10^{-7}B$ the complete chain of effective behavior is expected to go through the following sequence of effective regimes. At high $T$, the initial FO1 regime goes over into the SC1 regime which is then abruptly released back into a low temperature FO2 regime. Consequently, a cascade of low temperature local moment (LM2) regime followed by the final low temperature strongly-coupled (SC2) behavior can be observed. This can be attributed to a fully developed reentrant Kondo behavior and it indicates a possible appearance of two Kondo effects at two separated temperature regimes~\cite{Diniz-2020, Zalom-2021a} $T_1>\Delta/B$ (high temperature) and $T_2<\Delta/B$ (low temperature). However, the experimental observation of this effect for semiconducting TDOS function is challenged by extremely stringent conditions similar to the ones discussed in Ref.~\cite{Zalom-2021a} for SC-AIM as shown briefly in Sec.~\ref{subsec_reentrant_compar}.

Let us now demonstrate the temperature-dependent effective behavior  using the resulting thermodynamic properties shown in Figs.~\ref{fig_5}$(b)$, $(d)$ and $(f)$. As explained in Sec.~\ref{subsec_nrg}, particular degrees of freedom of the QD are effectively projected out in RG fixed points which becomes immediately translated into the corresponding thermodynamic properties. For example, by comparing the FO1 regime of Figs.~\ref{fig_5}$(e)$ and $(f)$ with the onset of the FO2 regime at $T/B \lesssim \Delta$ we see that the impurity entropy $S_\mathrm{imp}$ reaches the value of $\ln(4)$ at two very distinct temperatures. The QD can then be understood as an effective system of four states. Similarly, in LM2 the QD behaves effectively as two state system with single electronic occupation. In the end, QD may be effectively quenched to single level at  $T/B \gtrsim \Delta$ and then once again at $T/B \ll \Delta$ when sufficiently low $\Delta$ values are set as indicated in Fig.~\ref{fig_5}$(f)$. This is in-line with the effective behavior deduced from the flow of the rescaled $E_n$ values with the spin fluctuations $S_z^2$, represented in Figs.~\ref{fig_5}$(b)$, $(d)$ and $(f)$ by blue line, and it, therefore, supports the same conclusions. 


\begin{figure*}[ht]
	\includegraphics[width=2.00\columnwidth]{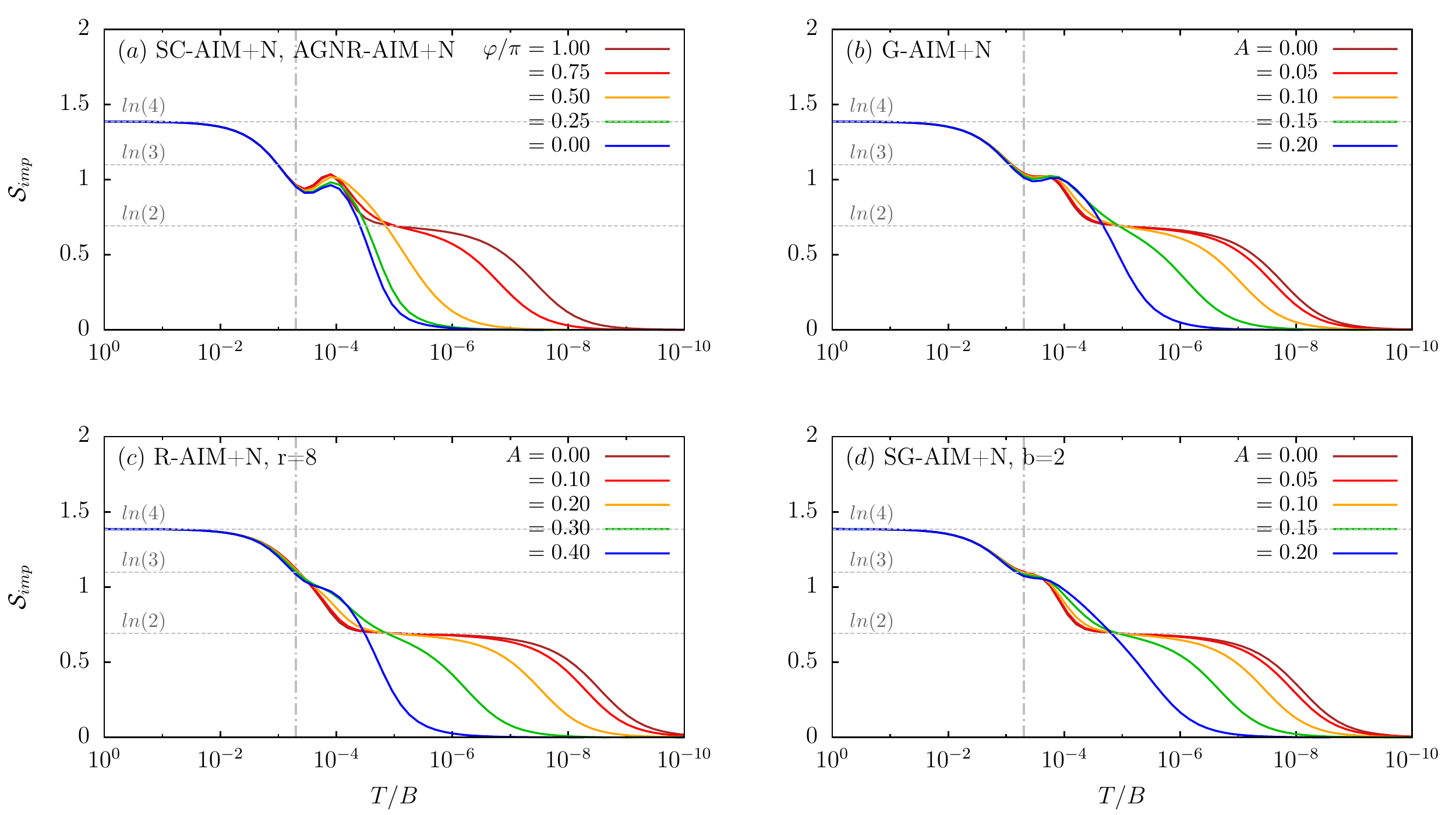}
	\caption{
		$(a)$
		Impurity entropy $S_\mathrm{imp}$ for SC-AIM+M and AGNR-AM+M model \eqref{DOS_AGNR} calculated at $U/\Delta=3$, $\Gamma_S/\Delta=1$, $\Gamma_M/\Delta=0.1$ and $\Delta=5 \times 10^{-4}B$ at indicated  \emph{ph}-asymmetry factors $\varphi$. Notice that $\varphi=\pi$ corresponds to fully symmetric case while $\varphi=0$ represents the largest possible asymmetry. 
		$(b)$
		Impurity entropy $S_\mathrm{imp}$ for G-AIM with model parameters according to panel $(a)$. Particle-hole asymmetry incorporated via factor $A$ with $A=0$ being the fully symmetric case while $A=1$ represents the largest possible asymmetry.
		$(c)$ 
		Impurity entropy $S_\mathrm{imp}$ for R-AIM according to \eqref{DOS_root} with $A$ defined as in panel $(b)$ and the same model parameters except of $\Gamma_S=6.25 \times 10^{-4}$ to adjust the overall TDOS integral. Note that large asymmetry factors $A$ are required to obtain comparable shifts in thermodynamic behavior since TDOS \eqref{DOS_root} essentially depopulates large regions around the gap edges. It thus implies that excess TDOS close to the gap edges, like in panel $(a)$, are essential for modifying the behavior of the QD system at low $T$.
		$(d)$
		Impurity entropy $S_\mathrm{imp}$ for SG-AIM with gap-edge singularities smoothed by cosine crossovers according to \eqref{DOS_cos}. Notice that quantitative differences to panel $(b)$ are almost negligible, ruling thus out the importance of finite discontinuities for qualitative behavior.
		\label{fig_7}
	}
\end{figure*}

\begin{figure*}[ht]
	\includegraphics[width=2.00\columnwidth]{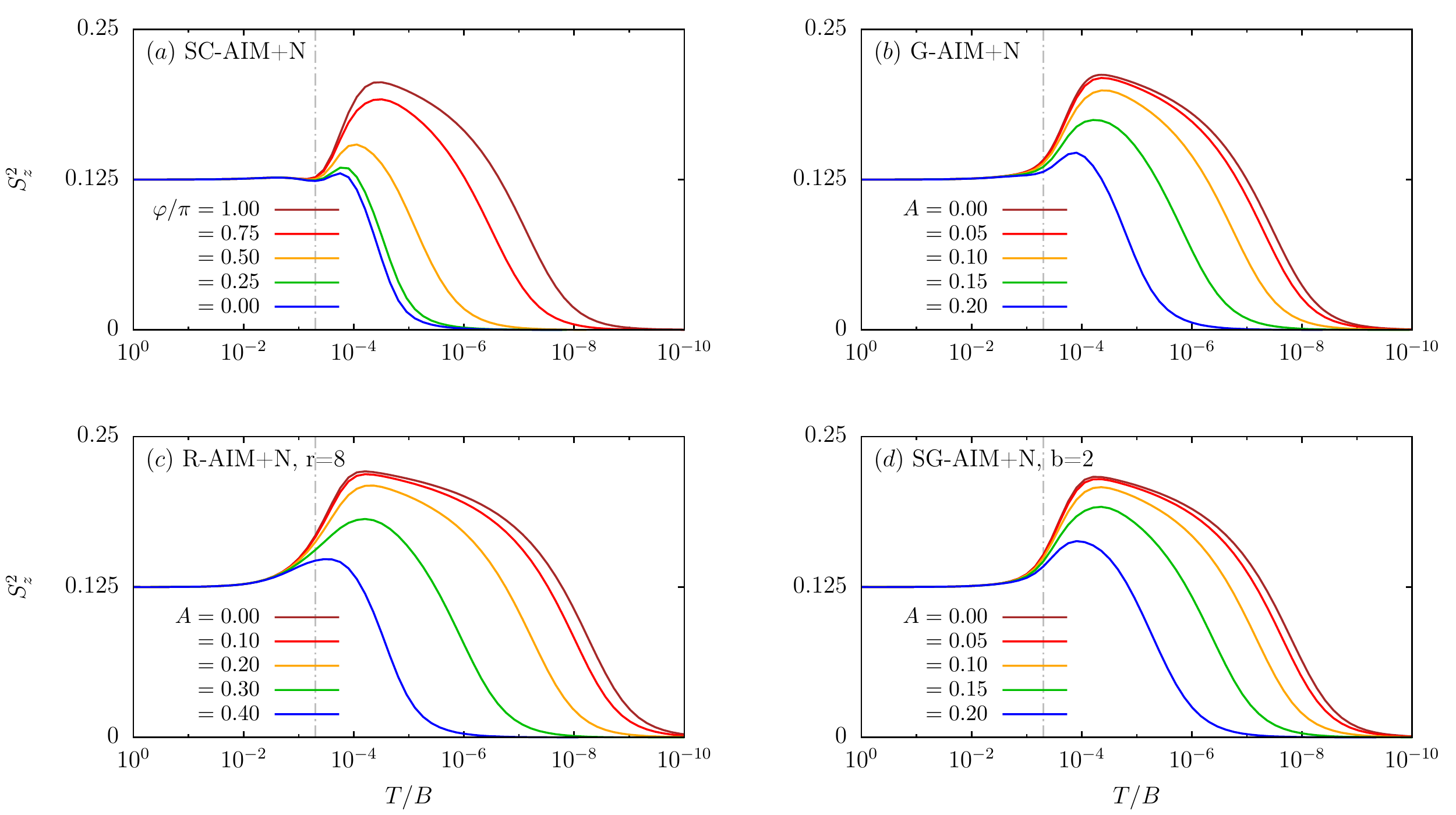}
	\caption{
		Panels $(a)$ -$(d)$ show NRG results for $S_z^2$ at parameter values corresponding to the same panels of Fig.~\ref{fig_7}.
		\label{fig_8}
	}
\end{figure*}

\subsection{Reentrant behavior of  \emph{ph}-asymmetric G-AIM \label{subsec_reentrant_asym}}

Above, we have established a clear equivalence of G-AIM with $A=0$ to the effective behavior of the half-filled AIM with flat TDOS. Here, we investigate case with finite \emph{ph}-asymmetry by selecting $A=0.1$. The resulting RG flow of rescaled eigenenergies $E_n$ is shown in Fig.~\ref{fig_6}. Once again, at  $n_\Delta\approx-2\ln\Delta/\ln\Lambda$ the flow separates into high and low temperature portions which, however, differ qualitatively from the $A=0$ case due to the presence of \emph{ph}-asymmetry in the TDOS. Most importantly, valence-fluctuation and frozen impurity regimes of AIM with out-of-half-filling QD appear which are accompanied by the corresponding values of $S_\mathrm{imp}$ and $S_z^2$. 

The resulting reentrant temperature behavior is thus marked by a cascade of effective models known from the AIM with out-of-half-filling QD. In the high temperature part, the free orbital regime FO1 emerges but the valence-fluctuation regime is skipped as before (due to the choice of $U$) and only at very small $\Delta$ values, see Fig.~\ref{fig_6}$(e)$, traces of the frozen impurity regime FI1 are visible. The effective projection of several levels of the QD is then abruptly lifted at $T \approx \Delta/B$ where the system can go either directly to the low temperature local-moment regime LM2 for large $\Delta$ as observed in Fig.~\ref{fig_6}$(a)$ or at moderate $\Delta$ values even to the low temperature free orbital regime FO2. However, at sufficiently small $\Delta$ values the low temperature frozen impurity regime FI2 is reached immediately, see Fig.~\ref{fig_6}$(e)$. The corresponding thermodynamic properties Figs.~\ref{fig_6}$(b)$, $(d)$ and $(f)$ and are in accord with temperature-dependent effective behavior that follows from the $E_n$ flow analysis. 

The deviation from the $A=0$ scenario are clearly dependent on the size of $\Delta$. In Figs.~\ref{fig_6}$(a)$ and $(b)$, at $\Delta = 5 \times 10^{-4}B$, we see negligible differences with the results in Figs.~\ref{fig_5}$(a)$ and $(b)$. Nevertheless, the flow of rescaled $E_n$ clearly shows energy splittings around the corresponding \emph{ph}-symmetric values --- a behavior that persists even to the smallest temperature scales. This requires a thorough explanation given the fact that the low temperature flow is intuitively expected to depend primarily on the gap region, where the metallic electrode induces constant and  \emph{ph}-symmetric TDOS. The same behavior was already observed in Ref.~\cite{Zalom-2021} for SC-AIM+M case and was clearly linked to the underlying $\Gamma_M=0$ version of the problem (SC-AIM). A detailed analysis of this issue is given at the end of this section.

The splittings induced by \emph{ph}-asymmetry in the supragap part of TDOS result in
diminishing difference between the ground state singlet and first excited doublet in the high temperature flow, as shown in Fig.~\ref{fig_6}$(e)$. Such behavior is a typical precursor for the singlet-doublet phase transition in $\Gamma_M=0$ systems. Note that such $E_n$ splittings could be noticed already for SC-AIM, see Fig.~(2) in Ref.~\cite{Hecht-2008}. However, the choice of parameters in the cited work obscured this observation. There chosen parameters seemingly lead to a conclusion that the RG flow of SC-AIM is similar to that of half-filled AIM. Yet, this is actually a consequence of fixing $\varphi=0$ and selecting only the extreme cases with $\Delta \gg T_K$ and $\Delta \ll T_K$, where $T_K$ is the Kondo temperature. For $\Delta \approx T_K$, which was not addressed in the cited work, the splitting of $E_n$ values, which points towards the correspondence to AIM with QD out-of-half-filling, would be clearly visible.

\begin{figure*}[ht]
	\includegraphics[width=2.00\columnwidth]{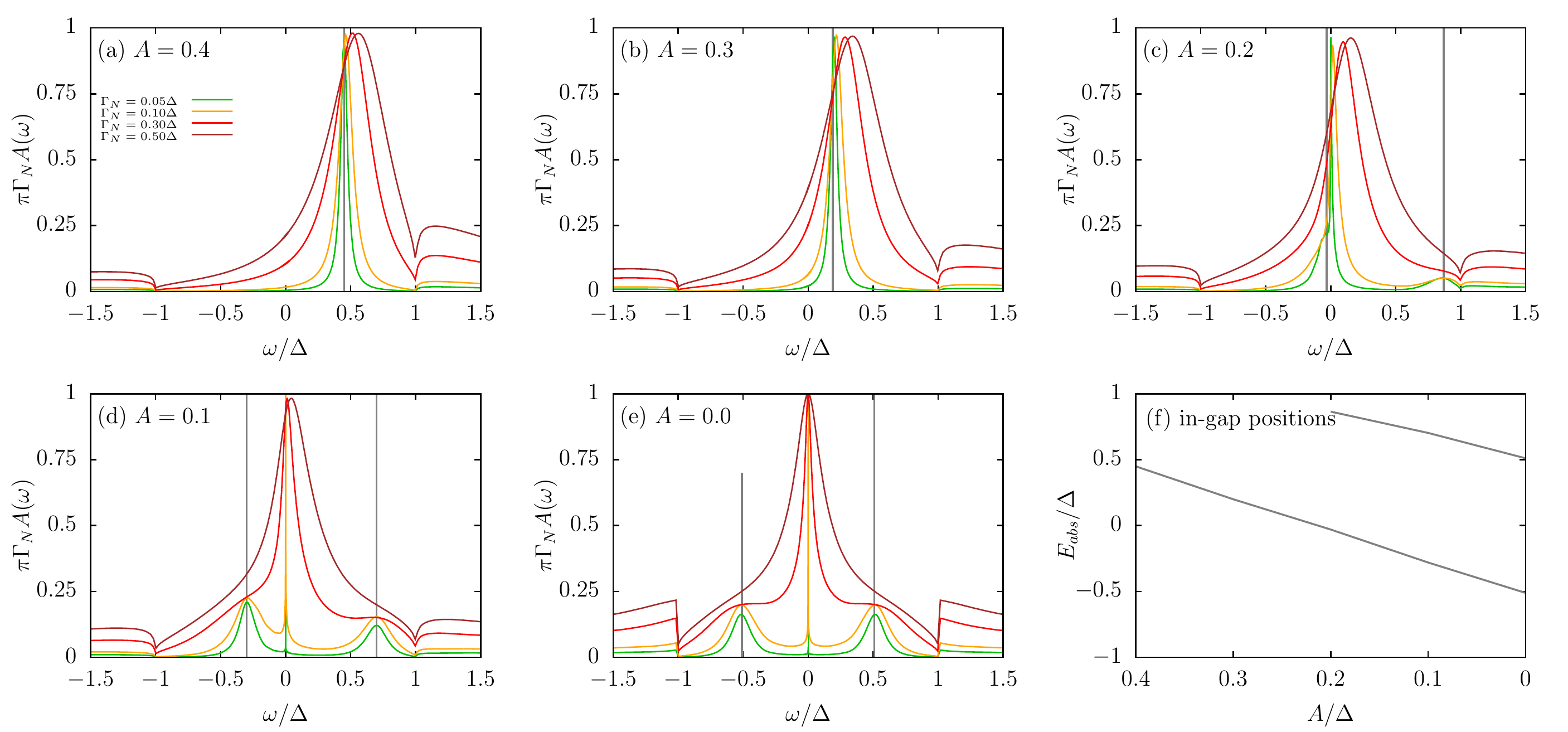}
	\caption{$(a)$-$(e)$ The in-gap part of the spectral functions shows broadened in-gap states of the hybrid normal-gapped reservoir with tunneling DOS (\ref{DOS_root})  for selected values of the asymmetry factor $A$ as a parametric plot of $\Gamma_M$. The vertical gray lines show the limiting positions of the expected position of sharp in-gap states for the $\Gamma_M \rightarrow 0$. $(f)$ $A$ dependence of sharp in-gap states of the gapped Anderson model (gray) and the position of maxima in their broadened analogs of the hybrid normal-gapped reservoir. All panels have been calculated using NRG Ljubljana with $\Lambda=2$ for $U=3\Delta$, $\Gamma_S=\Delta$, $\Delta= 10^{-3}B$ with $2B$ being the width of the band. \label{fig_9}
	}
\end{figure*}

To wrap it up, reentrant behavior critically hinges on two factors: the presence of well defined gap and \emph{ph}-asymmetry in the supragap region of the TDOS. The later ensures that the high temperature flow corresponds to AIM with correspondingly varying filling of the QD. Moreover, it also translates into the low temperature correspondence of the effective behavior to AIM with out-of-half-filling QD. In this respect, we note that  \emph{ph}-asymmetry introduced into the TDOS at supragap temperatures causes an effective behavior according to AIM with QD out of half-filling for all $T>\Delta$ but is most pronounced close to the threshold of $T \approx \Delta$. 
Because of the reentrant nature of the RG flow, the low temperature flow coupled to the metallic electrode starts from an effective model of non-half-filled QD. We can thus largely understand the situation as having an AIM with QD out-of-half-filling which couples to a metallic lead.


\subsection{Comparison of effective behavior for gapped host bands \label{subsec_reentrant_compar}}

Since thermodynamic properties are a direct consequence of the underlying effective models, we may streamline the comparison between the models by discussing just the temperature-dependent values of $S_\mathrm{imp}$ and $S_z^2$. The relevant results are presented in Figs.~\ref{fig_7} and \ref{fig_8}. In all presented cases we set  $U/\Delta=3$, $\Gamma_S/\Delta=1$, $\Gamma_M=0.1\Delta$ and $\Delta=5 \times 10^{-4}B$  and vary only the type and asymmetry of the TDOS.

In panels~\ref{fig_7}$(a)$ and \ref{fig_8}$(a)$ the thermodynamic properties of the SC-AIM+M and the AGNR-AIM+M model \eqref{BCS_DOS_w} are shown. Although, the underlying physics of these models is essentially different their effective behavior is exactly the same and follows the reentrant pattern discussed previously. The \emph{ph}-asymmetry factor $\varphi$ however governs only a quite narrow region around the gap. Thus, even at the highest induced \emph{ph}-asymmetry ($\varphi=0$) the ratio of the negative frequency portion of the TDOS to its positive frequency counterpart is only $\approx 1.008$ while the respective ration for asymmetry parameters $A$ in panels~\ref{fig_7}$(b)$-$(d)$, or~\ref{fig_8}$(b)$-$(d)$ is $(1+A)/(1-A)$ and thus much larger. 

Comparing first the \emph{ph}-symmetric results of SC-AIM+M and AGNR-AIM+M to the remaining three models, we notice that the Kondo temperature in Fig.~\ref{fig_7} might be read-off approximately as $T$ at which $S_\mathrm{imp}$ reaches $\ln(2)/2)$, which is sufficient for our purposes \footnote{More rigorously Kondo temperature should to be extracted from well converged spectral functions or magnetic susceptibility.}. We notice that Kondo temperature in Fig.~\ref{fig_7} is roughly ten times larger than in panels Fig.~\ref{fig_7}$(b)$-$(d)$. This clearly shows the importance of states close to the gap edges in screening processes of the QD. These findings are also confirmed by analogous comparison made for $S_z^2$ from Fig.~\ref{fig_8}. 

Incorporating \emph{ph}-asymmetry into the TDOS underscores these findings. As already pointed out, the overall asymmetry of the host band in Fig.~\ref{fig_7}$(a)$ is relatively small compared to Figs.~\ref{fig_7}$(b)$-$(d)$ and yet at $\varphi=0$ it still increased the transition to the strongly correlated regime by five orders of magnitude in comparison with $\varphi=\pi$. Comparable effects in the remaining three models are only possible at factors $A$ which induce very large \emph{ph}-asymmetry  into the TDOS. Although, this still offers a good theoretical tool for understanding the unified nature of QPTs the necessary large imbalance challenges the possible experimental realization as we discuss in more detail in Sec.~\ref{sec_exps}.

Comparison of panels $(b)$ and $(d)$ in Figs.~\ref{fig_7} and \ref{fig_8} respectively, shows that smoothening of finite discontinuities at the gap edges has overall small impact on the resulting effective behavior. Nevertheless, even here we notice that the reentrant behavior is actually weakened at $T\approx \Delta$ (dashed-dotted line). Similar findings also apply to the R-AIM case shown in Figs.~\ref{fig_7}$(c)$ and \ref{fig_8}$(c)$. Here the effect is even stronger because the shape of the TDOS causes redistributions of states in more extended regions. Consequently, much larger  \emph{ph}-asymmetry in the TDOS is required to obtain comparable effects on the thermodynamic behavior. Contrary, in AGNR-AIM+M the BCS-like singularities on the gap edges modify only regions very close to the gap edges but actually very large disproportions in positive negative frequencies are induced. We can thus conclude that modifications close to the gap edges play a more essential role in affecting the screening processes at $T \ll \Delta/B$ than the extended modifications present in SG-AIM and, especially, R-AIM.

\section{Underlying QPTs}


Since the metallic electrode added to the gapped TDOS functions dominates the low $T$ characteristics, the ground state properties for $\Gamma_M = 0$ can only be assessed indirectly. As shown in Figs.~\ref{fig_1} and \ref{fig_2} the subgap spectral functions can be obtained with a sufficient accuracy for AGNR-AIM and SC-AIM respectively. It not only allows to localize the (eventual) in-gap peaks but also to track their development with the change of \emph{ph}-asymmetry. We present such analysis in Sec.~\ref{subsec_spec_AIM} for G-AIM for different \emph{ph}-asymmetries. We confirm the presence of the in-gap states and also show their crossing with the Fermi energy as a response to \emph{ph}-asymmetry. An underlying QPTs in $\Gamma_M=0$ system must therefore be present. Next, we compare the evolution of extracted subgap peak positions for continuous TDOS functions of Sec.~\ref{subsec_spec_compar}. The limits of this approach as well as the possible departure from singlet-doublet related behavior are discussed in Sec.~\ref{subsec_escape} where continuous TDOS systems of SG-AIM and R-AIM are shown to allow an escape of in-gap states into the continuum.

\subsection{Subgap spectral features of G-AIM \label{subsec_spec_AIM} }

Due to the presence of the metallic lead standard NRG calculations could be carried on, including the determination of unbiased spectral functions. In Fig.~\ref{fig_9}$(a)$-$(f)$ we present the evolution of the subgap spectral functions of G-AIM+M with decreasing $\Gamma_M$ for various values of $A$ with vertical lines marking the extrapolated positions of the sharp in-gap states expected at $\Gamma_M = 0$. The panels of Fig.~\ref{fig_9} are ordered to match a corresponding evolution for SC-AIM or AGNR-AIM systems.

In detail, for the \emph{ph}-symmetric case, i.e., $A=0$, shown in Fig.~\ref{fig_9}$(e)$ one observes a symmetric subgap spectral function. The Kondo anomaly at $\omega=0$ is accompanied by two side peaks at $\omega/\Delta \approx 0.5$ visible for $\Gamma_M \lesssim 0.1 \Delta$. These cannot be understood as Hubbard peaks due to the mismatch in the position and traces of Hubbard-like satellites in the supragap region \footnote{We stress that Hubbard peaks due to bare $\Gamma_S$ and $U$ interfere with the high temperature traces of the Kondo peak, a situation analogous to that of Fig.~2.$(e)$ and $(f)$ in Ref.~\cite{Zalom-2021a}}. Increasing  \emph{ph}-asymmetry forces an excess TDOS in its negative (particle) domain which deforms the spectral function. For $A \lesssim 0.2$ the position of the Kondo peak stays fixed to $\omega=0$ and only the side peaks shift towards positive frequencies. However, at critical \emph{ph}-asymmetry, i.e., $A_c \approx 0.2$, the electron-like side-peak is pushed against the gap edge at $\omega=\Delta$ and its hole-like counterpart crosses the Fermi energy, see Fig.~\ref{fig_9}$(c)$. At this point the central Kondo peak is destroyed as a result of increased charge fluctuations in the system. In the terms of fixed points and effective models, this scenario is linked to reaching the effective FI regime by skipping VF and LM regimes. Increasing $A$ beyond  $A_c$ results in a further shift of the in-gap peak toward $\omega=\Delta$ position without any Kondo like maxima being present. The second in-gap peak vanishes soon and it is impossible to distinguish whether it traverses into the continuous part of the spectral function or merges with the gap edge at $\omega=\Delta$. The regime observed at $A<A_c$ is thus in full analogy to the $\pi$-like regime described in detail for SC-AIM+M, while for $A>A_c$ characteristic signs of $0$-like regime are evident \cite{Zalom-2021}.

\begin{figure*}[ht]
	\includegraphics[width=2.00\columnwidth]{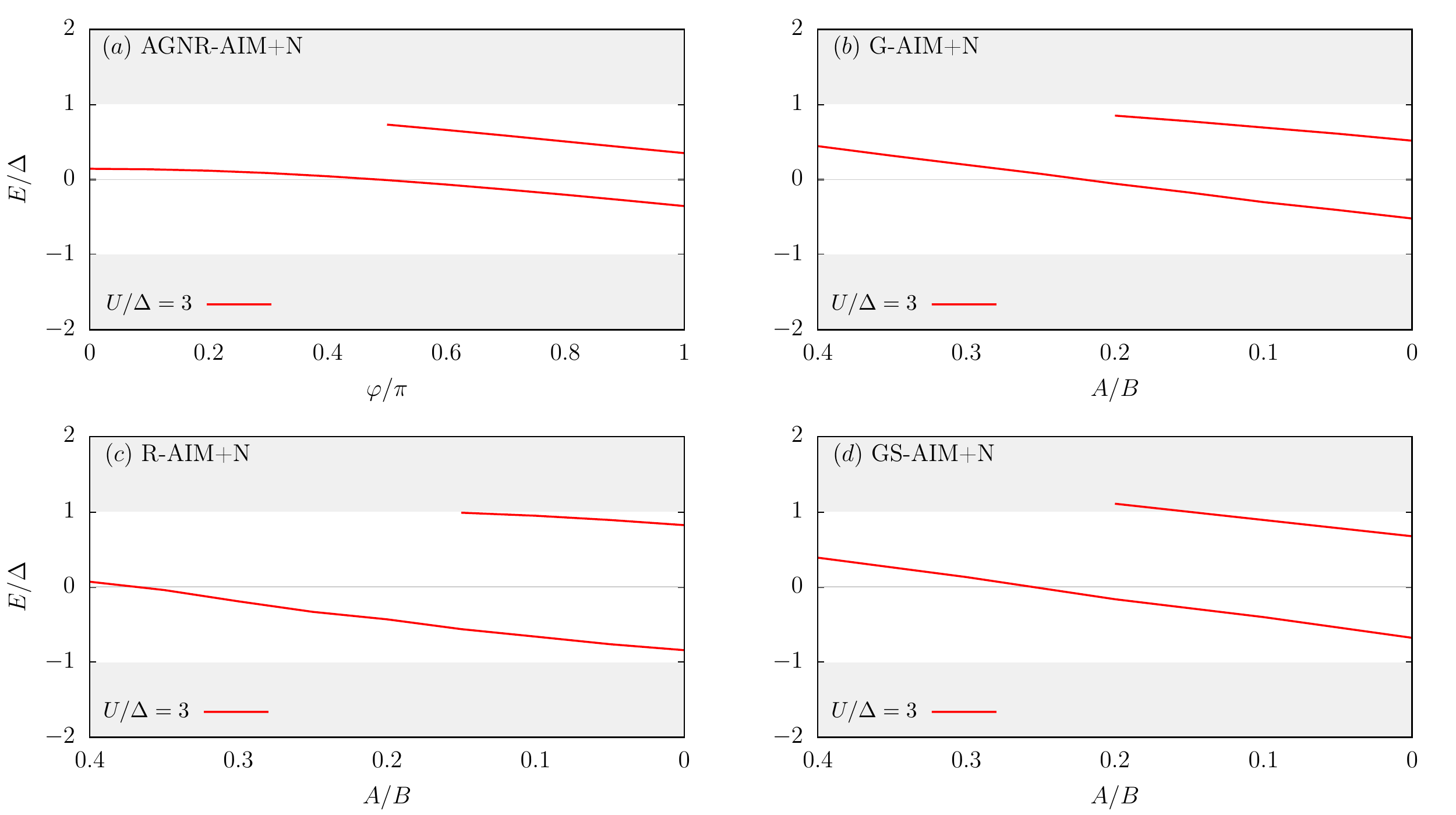}
	\caption{
		$(a)$
		Low lying states for SC-AIM and AGNR-AIM+M model \eqref{DOS_AGNR} calculated at $U/\Delta=3$, $\Gamma_S/\Delta=1$, $\Delta=5 \times 10^{-4}B$ and given $\Gamma_M$ in dependence of \emph{ph}-asymmetry factor $\varphi$. Notice that $\varphi=\pi$ corresponds to fully symmetric case while $\varphi=0$ represents the largest possible asymmetry. 
		$(b)$
		Low lying states for gapped AIM with model parameters according to panel $(a)$. Particle-hole asymmetry incorporated via factor $A$ with $A=0$ being the fully symmetric case while $A=1$ represents the largest possible asymmetry.
		$(c)$ 
		Low lying states for AIM with root-like bands according to \eqref{DOS_root} with $A$ defined as in panel $(b)$ and the same model parameters except of $\Gamma_S=6.25 \times 10^{-4}$ to adjust for the overall TDOS integral. Note that large asymmetry factors $A$ are required to obtain comparable shifts as TDOS \eqref{DOS_root} essentially depopulates large regions around the gap edges. Initially, no bound states are present as the center of the broadened peak lies out of the gap region. 
		$(d)$
		Low lying states for AIM with gap-edge singularities smoothed by cosine crossovers according to \eqref{DOS_cos}. Notice that quantitative differences to panel $(b)$ are almost negligible, ruling thus out the importance of finite discontinuities for qualitative behavior.
		\label{fig_10}
	}
\end{figure*}

Overall, we obtain subgap spectral behavior that is in complete analogy to the AGNR-AIM+M case presented in Fig.~\ref{fig_1} where the properties of the ground state at $\Gamma_M=0$ may be inferred from the corresponding SC-AIM solution. We can thus avoid the limitations of G-AIM by rationalizing the results of G-AIM+M in analogy to the AGNR-AIM+M case. Consequently, a doublet ground state at $A=0$ is determined which is, also, in accordance with results of various perturbative and non-perturbative results~\cite{Galpin-2008,Galpin-2008a,Galpin-2008,Galpin-2008a,Moca-2010,Chen-1998,Pinto-2015}. The crossing of the broadened in-gap peak with Fermi energy at $A_c/B=0.2$ indicates then an underlying QPT for $\Gamma_M=0$. Again, the nature of the ground state can not be read-off directly but comparison of Fig.~\ref{fig_9}$(f)$ with Fig.~\ref{fig_1}$(f)$, clearly points to a doublet-singlet ground state transition at $A_c$.

Note, that the number  of the sub-gap peaks and their asymmetric position with respect to Fermi level, as shown in Fig.~\ref{fig_9}$(f)$, seems to be in strong contrast with the SC-AIM case. However, the reason for the presence of symmetric pairs of ABS states in SC-AIM is the underlying formation of Cooper pairs ensuring that the energy eigenstates have half-to-half particle and hole character. Consequently, pairs of ABS states in spectral functions represent one-particle excitations of the ground state --- and this is reflected in the characteristic picture showing pairs of ABSs in the normal spectral function. Apart from this symmetry, which is directly linked to the superconducting nature of the host band, the QPT of G-AIM obtained by varying $A$ is the same as that of the generalized AGNR-AIM or SC-AIM case when QD is at half-filling.  In addition, although not explicitly presented here, the TDOS functions \eqref{DOS_cos} and \eqref{DOS_root} show similar behavior for the parameters used in Sec.~\ref{sec_thermo} and they can be addressed by the same qualitative arguments as applied to the G-AIM. 

In other words, a broad class of (non-superconducting) systems represented here by the cases \eqref{DOS_AGNR}, \eqref{DOS_flat}, \eqref{DOS_cos} and \eqref{DOS_root}, support only two distinct phases at $\Gamma_M=0$, namely the $0$ and the $\pi$ phase as known from SC-AIM. When metallic electrode is added, the distinct properties of the two phases are blurred-out and instead two regimes termed as $0$-like and $\pi$-like regimes appear \cite{Zalom-2021}. In detail, the $\pi$-like regime is for finite but small $G_N$ signaled by the appearance of central Kondo-like peak and two in-gap peaks. These are asymmetrically placed within the gap for any $A\neq0$,i.e., they do not form a symmetrized pairs known from the SC-AIM due to the reasons discussed above. When a critical value of \emph{ph}-asymmetry is reached, the charge-fluctuations take over and suppress the central peak. For models with weakly hybridized metallic electrode, one reaches the $0$-like regime which is marked by one in-gap peak being present in positive frequency region of the gap while none is present in the negative frequency gap region. A second peak is pushed toward the gap edge at $\omega=\Delta$ and, although, does not completely vanish for $\Gamma_M\neq 0$, it is highly suppressed and practically invisible in the background features. 
We also stress that the movement of the in-gap peaks toward positive frequencies is connected with how the \emph{ph}-asymmetry was incorporated into the TDOS. In the case of $0>A>-1$ we would observe the same behavior but in a reversed direction.

Let us conclude this section by noting that the above discussed critical $A$ is $U$ dependent. The relevant SC-AIM as well as SC-AIM+M studies~\cite{Zalom-2021,Zalom-2021a} imply that there is actually a threshold $U$ above which the $0$-like phase does not realize at all. The whole phase space is then dominated by the $\pi$-like regime and at $\Gamma_M=0$ a distinct $\pi$ phase is observed. Similarly, while qualitatively all systems behave essentially the same, variations of the TDOS functions can significantly influence particular details, e.g., the evolution of the subgap states. We briefly discuss this in the following section.

\subsection{Comparison of subgap spectral properties for gapped host bands \label{subsec_spec_compar} }

\begin{figure*}[ht]
	\includegraphics[width=2.00\columnwidth]{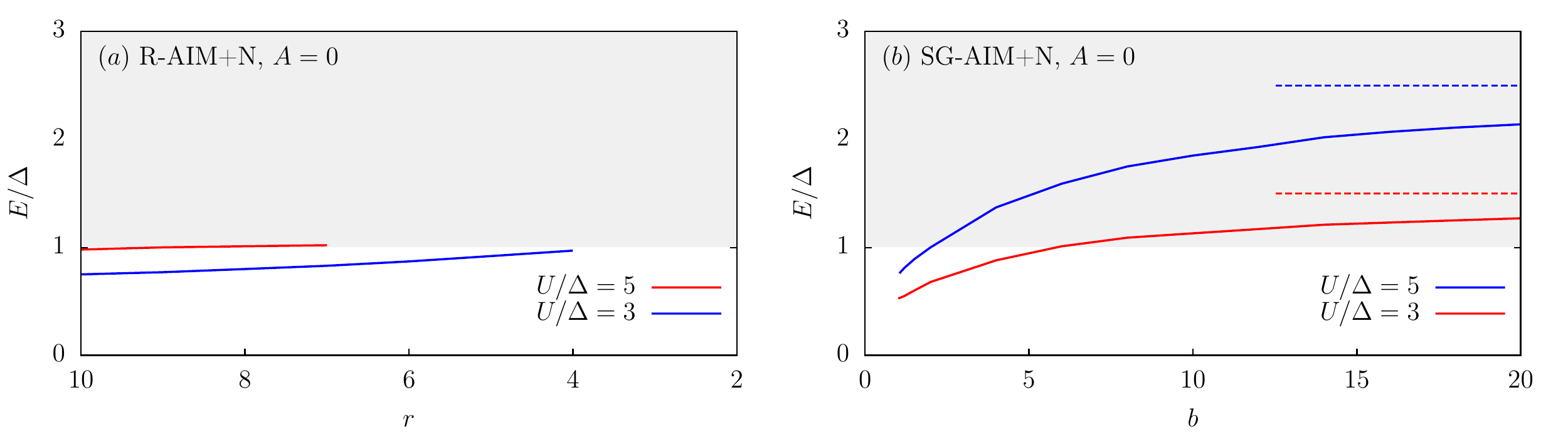}
	\caption{
		$(a)$
		Position of the inner most peaks for R-AIM with $A=0$ at two values of $U/\Delta$.
		$(b)$
		Position of the inner most peaks for SG-AIM with $A=0$ at two values of $U/\Delta$.
		\label{fig_11}
	}
\end{figure*}


Let us now compare the evolution of subgap peaks for the  investigated TDOS functions. To this end, we have extracted the subgap peak positions (if present) and plotted these in Figs.~\ref{fig_10} and~\ref{fig_11}. In all cases, the same qualitative dependence of subgap peak positions on the \emph{ph}-asymmetry, including the crossings of in-gap states with Fermi energy, is observed. Arguably, the crossing accompanied by the corresponding QPT might not happen always at experimentally meaningful values of $\varphi$ for AGNR-AIM and of $A$ for G-AIM, R-AIM and SG-AIM cases. Nevertheless, they still point toward the same singlet-doublet related properties of the TDOS systems under the study which was our intent to study. In Sec.~\ref{sec_exps}, we also discuss other, experimentally and physically more straightforward ways to induce QPTs into such systems.

For now, let us also briefly address the $U$-dependence of the in-gap peak positions (see Fig.~\ref{fig_11}) to point out some natural restrictions which are analogous to SC-AIM case. Without  \emph{ph}-asymmetry, e.g., at $A=0$ and $\varphi=\pi$, we observe that for $U=5\Delta$ the symmetric pair of in-gap peaks is pushed further apart compared to the $U=3\Delta$ case for all four cases (data shown for R-AIM and SG-AIM in Fig.~\ref{fig_11}) and all systems are in the doublet ground state. In this respect we note that from all TDOS cases studied here, the R-AIM case requires the highest amount of \emph{ph}-asymmetry to drive the system into QPT, as Fig.~\ref{fig_9} clearly demonstrates. Such values might not even be physically achievable. However, such a situation occurs also in SC-AIM where above certain threshold values of $U$ one might actually never induce enough \emph{ph}-asymmetry into the underlying TDOS after the mapping $\mathbb{T}$ based solely on tuning the phase drop $\varphi$ and the system remains then always in the $\pi$ phase. Nevertheless, we would like to stress that such scenarios are not generic and depend clearly in a multiparametric way on the exact realization. This applies to both superconducting and semiconducting host bands.

Let us also closely inspect the possibility of naive interpretation of the observed broadened in-gap peaks as Hubbard satellites. For $U/2>\Delta$ it is clearly ruled out by the presence of the reentrant temperature behavior in the presented cases. First, in the gap region ($|\omega|<\Delta$) all TDOS functions are actually \emph{ph}-symmetric, and indeed at approximately $\pm U/2$ the supragap spectral functions posses traces of Hubbard peaks which however interfere with the formation of the high temperature Kondo peak as described in \cite{Zalom-2021,Diniz-2020}. Additionally, also the RG flow and resulting thermodynamic effective behavior established that in all four general cases reentrant behavior appears which is linked with the presence of two scales into the system. One is linked with high $T$ behavior and with the ordinary Hubbard peaks (deformed due to the presence of \emph{ph}-asymmetry). On the other hand, the low $T$ behavior leads to the formation of broadened in-gap peaks which can be be formally understood as effective Hubbard peaks but as shown in Ref.~\cite{Zalom-2021a} they do not follow from the bare values of $\Gamma_M$ and $U$ and are heavily influenced by supragap properties which govern their position.

However, let us note that for $U/2<\Delta$ our current approach is insufficient and a dedicated NRG solution for $\Gamma_M=0$ is required to give a precise answer. The reason lies in the fact that for such a choice of $U$ the Hubbard peaks are not present in supragap portion of the spectral function, the reentrant character of the flow is very weak and charge oscillations influence directly the in-gap spectral function with in-gap peaks becoming indistinguishable from \emph{ph}-asymmetry dependent Hubbard peaks. They might therefore be just an artifact of augmenting the gapped TDOS functions by metallic leads and the true in-gap peaks (if present) are either hidden behind their strong signal, or behind the Kondo anomaly as for small $U$ any in-gap peak is expected to lie close to the Fermi energy.

\subsection{Escape of broadened in-gap peaks into the continuum\label{subsec_escape}}

The SG-AIM and R-AIM have a continuous TDOS profile at the gap edges and leakage of in-gap peaks into the continuous supragap spectrum might therefore be expected. Thus, when broadened in-gap peaks are observed outside of the gap region they cannot be strictly bound for the $\Gamma_M=0$ counterpart of SG-AIM+M or R-AIM+M. On the other hand G-AIM is defined to always posses a finite discontinuity which is expected to provide a sufficient barrier towards any leakage. However, AGNR-AIM and the equivalent SC-AIM cases after the transformation $\mathbb{T}$ posses some continuous features that require a brief discussion. We stress that for any $\varphi \neq 0$ BCS-like divergences appear at the gap edges $\omega \pm \Delta$. However at $\varphi = 0$ the divergence at $\omega=\Delta/B$ vanishes and the profile becomes proportional to $|\omega|\sqrt{\omega-\Delta}$ for $\omega>\Delta$, which is similar to R-AIM profile. Nevertheless, no leakage of the inner in-gap state was reported here. This is probably a consequence of having yet another BCS-like divergence at $\omega=-\Delta/B$. We therefore focus on continuous TDOS cases of R-AIM and SG-AIM here.

First, let us analyze the $A=0$ R-AIM system with its in-gap peak positions shown in Fig.~\ref{fig_11}$(a)$. Here, the shaded region shows the supragap portion of the spectra while below $E/\Delta<1$ the broadened in-gap peaks are observed (the Kondo anomaly at $\omega=0$ is omitted). Negative frequency regions are symmetric to the region shown in Fig.~\ref{fig_11}$(a)$. Two values of $U/\Delta$ are used in order to demonstrate the expelling of broadened in-gap peaks (if present in the gap) to the gap edges with enhanced $U$. For R-AIM+M with $r=10$ we observe a well separated broadened in-gap state at $\omega/\Delta \approx 0.8$ for $U/\Delta=3$. However, already at $U/\Delta=5$ it becomes hardly distinguishable with the supragap part of the spectral function as it is expelled onto the gap edge $\omega/\Delta=1$. 

For R-AIM systems with $r$ in the interval of $10 \leq r \leq 4$ it appears that an in-gap peak might be present for $U/\Delta=3$ and lie in the proximity of the gap edge. For $r\lesssim 4$ no features could be reliably identified in supra- as well as subgap parts of the spectral functions for $U/\Delta=3$. For $U/\Delta=5$ the situation is analogous but is observed already at $r \approx 4$ and there are two reasonable ways how to interpret such findings. The first is that the in-gap states for R-AIM do exist for given threshold value of $r^*$ until they reach the gap edges. The excited states with exactly the same quantum states as before might exist also for $r>r^*$ but obtain resonant character as they move into the continuum. The second possibility is, that they are not present at all which would break the analogy to the SC-AIM here.

Our present approach cannot give a conclusive answer which interpretation is correct. Nevertheless, we stress that even when it will be the later one, other here discussed analogies with AGNR-AIM+M and SC-AIM+M systems will be preserved. These include the presence of subgap Kondo peaks and traces of Kondo peaks in supragap spectral function. 

In SG-AIM+M systems the behavior becomes even more interesting as shown in Fig.~\ref{fig_11}$(b)$. Note that by setting $b=1$ we completely recover the G-AIM+M system and its corresponding broadened in-gap peaks. As before, increasing $U$ by fixing all other model parameters shifts the in-gap peak against the gap edge $\omega/\Delta=1$ which is a behavior analogous to that of AGNR-AIM or SC-AIM (where additional symmetrizations need to considered however). Increasing $b$ makes the discontinuity of G-AIM+M smoother which can also be understood as effective broadening of the gap region. Shifting of the in-gap states towards the gap edges is than an expected result. However, unlike in R-AIM+M systems once the broadened in-gap peaks reach the smooth gap edge they do not vanish and we may observe their transition into the supragap region. Since at $\Gamma_M=0$, there is non-zero TDOS in this region they cannot have bound character and are thus part of the continuous spectrum. Nevertheless, this behavior can still be attributed to the effective expansion of the gap region due to $b$.

The positions of the peaks is approaching approximately the $\pm U/2$ positions which correspond to the positions of Hubbard peaks. It is thus obvious that increasing the smoothness of the gap, their properties trivialize to those of ordinary Hubbard peaks. Incorporating  \emph{ph}-asymmetry into the TDOS will then force their movement towards Fermi energy. In such a case, it is difficult to make any conclusions about the correspondence to AGNR-AIM or SC-AIM cases. Consequently, it would be highly speculative to confirm or exclude the presence of singlet-doublet transitions or any other QPTs. We stress that the escape into the continuum is also accompanied by weakening of reentrant behavior described in Sec.~\ref{sec_thermo} and only underlines the importance of careful RG analysis. 

\section{Possible experimental realization of the singlet-doublet transition in semiconducting host bands \label{sec_exps} }

In this paper, we have postulated several shapes of TDOS functions and have been able to verify for all, 
that they support the presence of singlet-doublet QPTs. The existence of QPTs hinges more on the presence of spectral gap rather than on to the exact details of the TDOS. Consequently, we expect that quite a large class of set-ups (some already under investigation) might be promising candidates for experimental scrutiny of hereby presented results. 

In this respect, nanostructure, e.g., a single molecule or a carbon nanotube, absorbed on (or connected to) a semiconducting surface represent a generic layout~\cite{li2017anomalous}. In principle, the subgap states should be directly detectable by high resolution energy spectroscopy using tunneling probes. These techniques, besides their high precision, allow for some control of the heterostructure parameters, such as the hybridization or level energy~\cite{Temirov:Nanotech08,oberg13,Zonda-2021}. In practice, intraatomic interactions between the nanosystem and the substrate(s) might interfere with the simple single spin-orbital picture used in the present paper. As an example, one can mention the experiment in Ref.~\cite{li2017anomalous} where a layer of semiconducting graphene on a metallic substrate was studied. This seems to be an ideal realization od here presented model systems. However, the electronic transparency of graphene caused the nanosystem to develop strong spin interactions with the metallic substrate. Consequently, the resulting system did not inherit many of the semiconducting properties of graphene and any traces of possible QPTs as discussed in the present paper have been lost. Nevertheless, we stress that such effects are not of fundamental nature and do not hinder electron spectroscopy in semiconductor devices. As already proven for the ABS states in various types of Josephson junctions, high resolution scanning tunneling images might be even acquired~\cite{Chang-2013,Pillet-2010,Pillet-2013,Delagrange-2016}.  

Let us therefore focus on the fundamental constraint given by the ratio of $U/\Delta$. As discussed above and well known from superconducting realizations, a moderate value is required. In the case of strong interaction $U \gg \Delta$, the subgap states are pushed to the edge of the gap where they become indistinguishable with the continuum since all experiments are done at finite temperatures. In the other limit, $U \rightarrow 0$, they move to the close vicinity of the Fermi level, where they overlap and might even be mistaken for, or can merge with, other types of resonances, e.g., the Kondo anomaly (or Majorana states). This would again, complicate the observation of the Fermi-level crossing. 

However, semiconductors typically have larger values of the gap parameter and for the present study we require very clean samples since any dopants could interfere with the subgap spectrum. Thus, relatively large absolute values of $U$ are needed. In the case of superconducting quantum dot realizations, for which singlet-doublet QPTs have been already studied experimentally, the typical charging energy on the dot, made for example from the carbon nanotube, is in the order of $meV$ and the gap is in the order of hundreds of $\mu EV$ (e.g., $0.17$~meV for aluminum leads)~\cite{Maurand-2012,Delagrange-2015,Delagrange-2016}. 

Nevertheless, principally even semiconductors with very narrow gaps can be synthesized. For example mercury-cadmium telluride, used in infrared detectors, allows to tune the bandgap from zero up to few electron volts~\cite{hansen1982}. Heterostuctures consisting of organic molecules absorbed on the semiconducting surface, might also represent good alternatives. For example, the recent realization of porphyrin-based magnetic molecule TBrPP-Co on AGNR substrate~\cite{li2017anomalous} has $U$ in the order of few $eV$ while the AGNR band gap was $1-2eV$ in size depending on the width of the absorbed molecule. 

To our knowledge, no fundamental obstacles are present for detecting subgap states on top of semiconductor surfaces. However, unlike in superconductors where the phase bias controls the underlying \emph{ph}-asymmetry of the $w$ Bogolyubons, one cannot exert such precise tunning in semiconductors. Some materials, like AGNR, for example, allow to adjust the gap by external electric fields but to tune \emph{ph} asymmetry of the whole band might be extremely difficult. We therefore stress that the parameter $A$ used in this paper was solely used to theoretically model the singlet-doublet QPTs in a close analogy to superconducting realizations. Experimentally, more viable concept is readily available by using gate voltage on the QD. Driving it out-of-half filling, is then an equivalent process to the inclusion of \emph{ph}-asymmetry into the TDOS. Other alternatives to be considered include changing the size of the gap by tuning of the substrate or varying hybridization by lifting the molecule via STM tip. Experiments with carbon nanotubes seem to offer another pathway for detection of singlet-doublet QPTs in semiconductor host bands. To this end one may electrostatically control the the coupling of the carbon nanotube to the leads. Increasing the hybridization of the nanotube to the semiconductor would bring the system out of the doublet into the singlet ground state as already performed in superconductor QD devices \cite{Saldana-2020}.

%
%
%
%

\section{Conclusions \label{sec_conc}}

In this paper, we have analyzed half-filled QD coupled to four general classes of semiconductor TDOS functions (see Fig.~\ref{fig_0}) in regard to the existence of eventual singlet-doublet QPTs which are well established for superconducting host bands. 
To allow for an unbiased and standard NRG approach without temperature truncations, as employed in Refs.~\cite{Chen-1998,Moca-2010}, we have augmented the gapped host bands by a weakly hybridized metallic electrode. 
The thermodynamic behavior conformed to the underlying cascade of reentrant effective models that develop in two sequences. One is present at high and the other at low temperatures. Each sequence of effective models was established to follow those of AIM with out-of-half-filling QD albeit we explicitly demanded half-filled QDs in all systems studied here. 

Since metallic electrons enforce a singlet ground state into the system, we have observed only broadened in-gap features at $\Gamma_M \neq 0$ and the nature of QPTs could only be deduced indirectly. However, we stress that all of this limitations have been compensated by obtaining unbiased and numerically exact quantities. 

\begin{figure}[t]
	\includegraphics[width=1.00\columnwidth]{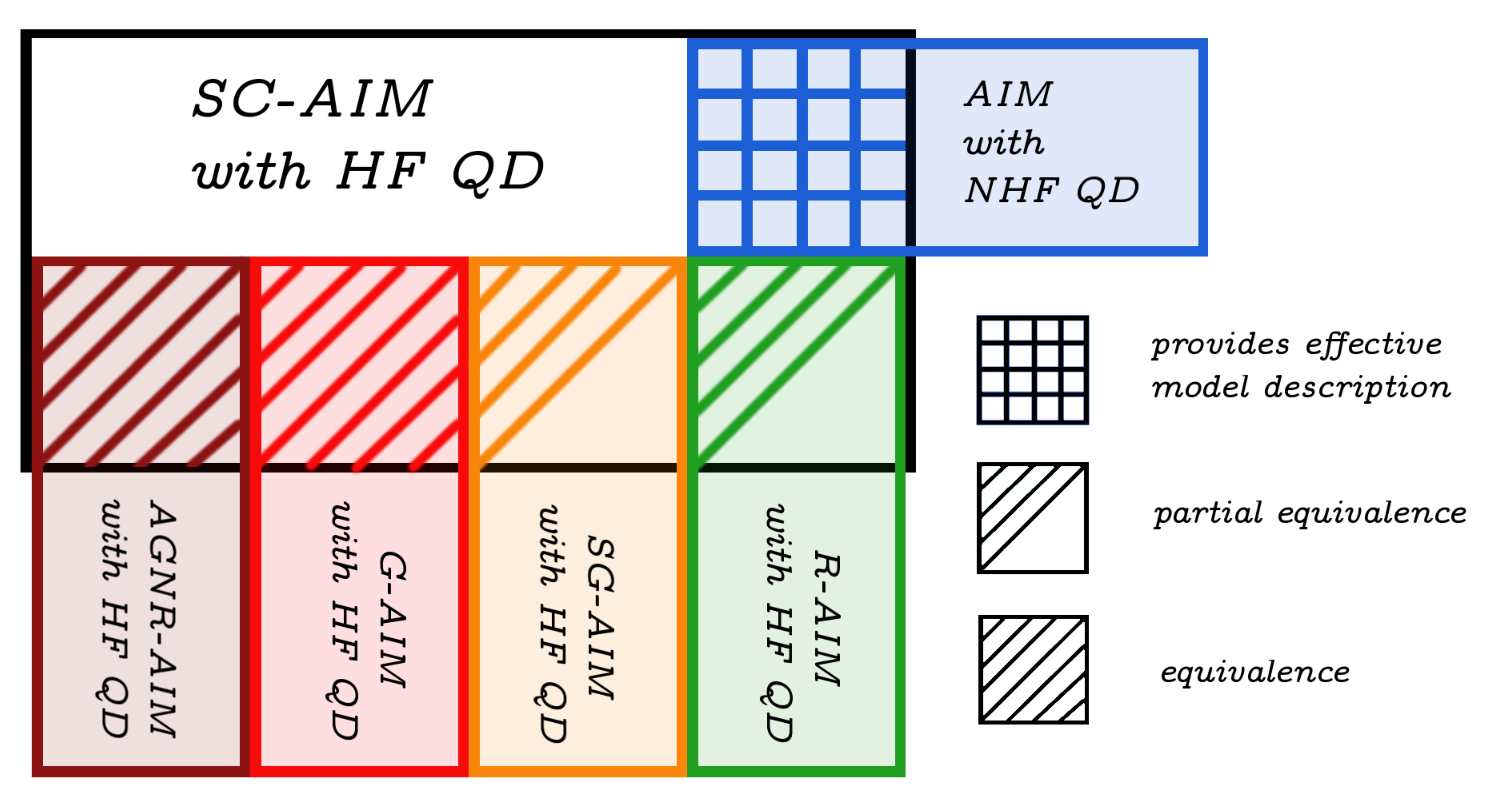}
	\caption{
		The main conclusion regarding the presence of singlet-doublet QPT in QDs coupled to gapped host bands of various properties is summarized as overlaps between the colored boxes representing a given model, i.e., AGNR-AIM with half-filled (HF) QD (brown), G-AIM with HF QD (red), SG-AIM with HF QD (orange) and R-AIM with HF QD (green), and the large box representing SC-AIM (black). AIM with out-of-half-filling (NHF) QD (blue box) provides an effective description. For R-AIM and SG-AIM we have shown in Sec.~\ref{subsec_escape} that our present approach with augmenting the host band with metallic electrode to perform standard NRG cannot reliably draw conclusion in some parameter regimes. 
		\label{fig_12}
		}	
\end{figure}

In general all four investigated classes may support the existence of singlet-doublet QPTs known from SC-AIM. Their physical origin lies in the presence of a gap and \emph{ph}-asymmetry in the TDOS. Our main conclusion can thus be summarized by Fig.~\ref{fig_12} in the following way. A prototypical SC-AIM case and all four general classes of semiconductor support the presence of singlet-doublet behavior. The spectral equivalence is strong for the G-AIM and AGNR-AIM systems since these posses discontinuities at the gap edges. However, once the discontinuities are smoothed-out, as in the R-AIM and SG-AIM cases, the equivalence becomes weaker. In-gap peaks known from SC-AIM may escape into the continuum as shown in Sec.~\ref{subsec_escape} and their properties become increasingly similar to the ordinary Hubbard peaks. In such cases, we can not confirm or exclude the presence of QPTs in such systems as the nature of the ground states can not be reliably extrapolated from model augmented by metallic electrodes. Further investigation therefore requires a genuine RG approach to gapped host bands which is currently being developed. In Fig.~\ref{fig_12} we indicate this by stating only a partial equivalence.

Crucially, the presence of in-gap states in any of the gapped systems in their realization with weakly hybridized leads is always signaled by reentrant effective-temperature behavior. This is governed by AIM with flat density of states but an out-of-half-filling QD. As explained in Sec.~\ref{subsec_reentrant_asym}, this is expected in high temperature part of the reentrant behavior which is indeed marked by the presence of  \emph{ph}-asymmetry in the supragap part of the TDOS. In the low temperature part a sequential nature of the flow can be interpreted as adding a flat metallic TDOS to the effective model of AIM with QD out-of-half-filling. Reentrant effective behavior follows therefore a cascade of effective models known from AIM with QD out-of-half-filling. This is a precursor for the newly introduced concept of reentrant Kondo behavior \cite{Diniz-2020,Zalom-2021,Zalom-2021a} but also for the appearance of singlet-doublet QPTs. Such behavior is analyzed in detail in Figs.~\ref{fig_5}-\ref{fig_8}.

We stress that in the present paper \emph{ph}-asymmetry was introduced to serve a theoretical purpose of showing that doublet-singlet QPTs exist in general classes of gapped host bands and are analogous to the well understood effects in superconducting realizations. The \emph{ph}-asymmetry was thus introduced into the TDOS to match its generation in SC-AIM case via tunning of the phase drop between two superconducting leads which, as shown in Ref.~\cite{Zalom-2021}, controls the \emph{ph}-asymmetry in the underlying description via Bogolyubons $w$. In semiconducting bands, an analogous way would demand to tune \emph{ph}-asymmetry in host bands which is practically hard to achieve. Second, the required \emph{ph}-asymmetry to force the resulting system out of the doublet ground state is large and therefore of limited experimental use. Instead, as proposed in Sec.~\ref{sec_exps}, an experimental observation should rely on taking the QDs out of their half-filled point which can be done by using appropriate gating. We have however deliberately singled out this scenario to future calculations since the corresponding phase space is large and the shear amount of data would make the presentation of the main findings difficult. We also stress that the obtained data are of importance for benchmarking the NRG algorithm for gapped host bands, e.g., SC-AIM after the transformation $\mathbb{T}$, which we are currently developing.

\begin{acknowledgements}  
	This work was supported by Grant No. 19-13525S of the Czech Science Foundation (M\v{Z}) and by the COST Action NANOCOHYBRI (CA16218) (PZ). 
	PZ was partially supported by Operational Program Research, Development and Education(financed by European Structural and Investment Funds and by the Czech Ministry of Education, Youth, and Sports), Project No. SOLID21-CZ.02.1.01/0.0/0.0/16\_019/0000760). The authors thank Pascu Moca, Vladislav Pokorn\'{y} and Tom\'{a}\v{s} Novotn\'y for helpful discussions. 
\end{acknowledgements}

\end{document}